\documentclass[journal]{IEEEtran}
\usepackage[underline=false]{pgf-umlsd}
\usepackage[linesnumbered, ruled]{algorithm2e}
\usepackage{amsmath,amssymb,amsfonts}
\usepackage{graphicx}
\usepackage{graphics}
\usepackage{textcomp}
\usepackage{xcolor}
\usepackage{tikz}
\usetikzlibrary{fit, calc}
\usepackage{pgfplots}
\usepackage{comment}
\usepackage{gensymb}
\usepackage[normalem]{ulem}
\usepackage{footnote}

\usepackage[flushleft]{threeparttable}
\usepackage{bbm}
\usepackage{amsthm}
\usepackage[english]{babel}
\SetAlCapNameFnt{\normalsize}
\SetAlCapFnt{\normalsize}
\usepackage[subrefformat=parens,labelformat=parens,caption=false,font=footnotesize]{subfig}

\graphicspath{{figures/}} %.eps figures

\def\BibTeX{{\rm B\kern-.05em{\sc i\kern-.025em b}\kern-.08em
       T\kern-.1667em\lower.7ex\hbox{E}\kern-.125emX}}

\newtheorem{definition}{Definition}

\makesavenoteenv{tabular}
\makesavenoteenv{table}

\pgfplotsset{compat=newest}

\newcommand*{\tikzmk}[1]{\tikz[remember picture,overlay,] \node (#1) {};\ignorespaces}
\newcommand{\boxit}[3]{\tikz[remember picture,overlay]{\node[yshift=3pt,fill=#1,opacity=.1,fit={(#2)( $(#3)+(.89\linewidth,.8\baselineskip)$ )}] {};}\ignorespaces}

\newcommand{\etal}{\textit{et al. }}
\newcommand{\fig}[1]{Fig. \ref{#1}}
\newcommand{\figs}[2]{Fig. \ref{#1} and \ref{#2}}
\newcommand{\figss}[3]{Fig. \ref{#1}, \ref{#2}, and \ref{#3}}
\newcommand{\apx}[1]{Appendix \ref{#1}}
\newcommand{\card}[1]{\mathrm{card}(#1)}
\newcommand{\argmax}{\mathrm{argmax}}
\newcommand{\too}[1]{\mathcal{O}\left(#1\right)}

\def\figscale{0.75}

\newcommand{\titleheader}{Accepted in IEEE Transactions on Wireless Communications}

\makeatletter
\def\ps@headings{%
\def\@oddhead{\mbox{}\scriptsize \titleheader \rightmark \hfil \thepage}
\def\@oddfoot{\scriptsize \@date \hfil }%
}

\def\ps@IEEEtitlepagestyle{%
\def\@oddhead{\mbox{}\scriptsize \titleheader \rightmark \hfil \thepage}%
\def\@oddfoot{\scriptsize \@date \hfil }%
}
\makeatother

\begin{document}

\title{Multi-Agent Reinforcement Learning for Adaptive User Association in Dynamic mmWave Networks}

\author{\IEEEauthorblockN{Mohamed Sana\thanks{Part of this work has been presented in IEEE Globecom 2019 \cite{sana2019UA}.}\thanks{Mohamed Sana and Emilio Calvanese Strinati are with CEA-LETI Minatec Campus, 17 rue des Martyrs, 38054 Grenoble, France (e-mail: mohamed.sana@cea.fr, emilio.calvanese-strinati@cea.fr).}, Antonio  De Domenico\thanks{Antonio De Domenico was with CEA-LETI when this manuscript was submitted. He is now with Huawei Technologies, Paris Research Center, 20 quai du Point du Jour, Boulogne Billancourt, France (e-mail: antonio.de.domenico@huawei.com).}, Wei Yu, Yves Lostanlen\thanks{
Wei Yu and Yves Lostanlen are with the  Department of Electrical and Computer Engineering, University of Toronto, Toronto ON, Canada M5S 3G4. Wei Yu is supported by the Natural Sciences and Engineering Research Council of Canada (e-mail: weiyu@ece.utoronto.ca, yves.lostanlen@ieee.org).}, and Emilio Calvanese Strinati}}
\maketitle

\maketitle

\pagestyle{headings}
\thispagestyle{empty}

\begin{abstract}
Network densification and millimeter-wave technologies are key enablers to fulfill the capacity and data rate requirements of the fifth generation (5G) of mobile networks. In this context, designing low-complexity policies with local observations, yet able to adapt the user association with respect to the global network state and to the network dynamics is a challenge. In fact, the frameworks proposed in literature require continuous access to global network information and to recompute the association when the radio environment changes. With the complexity associated to such an approach, these solutions are not well suited to dense 5G networks. In this paper, we address this issue by designing a scalable and flexible algorithm for user association based on multi-agent reinforcement learning. In this approach, users act as independent agents that, based on their local observations only, learn to autonomously coordinate their actions in order to optimize the network sum-rate. Since there is no direct information exchange among the agents, we also limit the signaling overhead. Simulation results show that the proposed algorithm is able to adapt to (fast) changes of radio environment, thus providing large sum-rate gain in comparison to state-of-the-art solutions. 
\end{abstract}

%\begin{IEEEkeywords}
%Cognitive Radio Network, Distributed User Association, Deep Multi-Agent Reinforcement Learning, millimeter-wave communications, micro-wave and millimeter-wave integration.
%\end{IEEEkeywords}

\section{Introduction}
The fifth generation of mobile communications (5G) promises to bring an unprecedented improvement in current wireless systems, providing ultra-high speed communications, extremely low latency, and enabling a plethora of use-cases. One of the main targets of 5G is to provide enhanced mobile broadband (eMBB) services, which are characterized by high data throughput requirements, e.g., the 5G target for downlink peak data rate is 20 Gbps \cite{TR38.913}. To boost the network capacity, 5G envisions the use of millimeter-wave (mmWave) bands, which offer large spectrum resources, together with the dense deployment of small cells in the network architecture. However, mmWave transmissions suffer from severe path-losses and are highly sensitive to blockages \cite{DeDomenico2017}. Moreover, network capacity does not increase systematically with the densification of base stations (BSs) due to e.g., channel interference and inefficient resource allocation. In this context, optimally associating user equipments (UEs) to BSs is a fundamental task to take full advantage of network densification and mmWave technology. This task is particularly challenging in dense 5G networks, and it is even more complex when taking into account channel interference, fast fading, and network traffic, which we later refer to as the environment dynamics. So far, the current solutions proposed in the literature typically require global information of the network, are computationally expensive, and do not consider the dynamic nature of wireless networks. More specifically, either the state-of-the-art algorithms are re-computed periodically or whenever a notable change has occurred in the environment to correct possible drifts from the optimal association. With the complexity carried by such approaches, they may lead to significant computation and signaling overhead. Therefore, there is the need for flexible and adaptive solutions with respect to environment dynamics.

In this paper, we address this issue by proposing a distributed algorithm based on multi-agent reinforcement learning where each UE acts as an independent agent that operates in a fully distributed manner. UEs learn by experience, based only on their local and partial observations in order to maximize the network sum-rate, as we focus on mmWave technology for eMBB services. Specifically, UEs learn to map their observations of the environment to actions that correspond to connection requests to their surrounding BSs. Our proposed framework has the advantage of being able to incorporate the environment dynamics during the learning phase i.e., the time varying nature of mmWaves channels and the traffic of each user, so that the user association is self-reorganized toward the optimal association when a relevant change occurs in the environment. Furthermore, we reduce the signaling overhead since agents does not exchange any direct information. Moreover, a salient feature of the proposed mechanism is that it can easily be leveraged to optimize operation in other wireless systems e.g., WiFi, without additional complexity. Eventually, the distributed nature of our algorithm alleviates the computation overhead, thus making the proposed framework a promising tool for next generation of wireless systems \cite{emilio2019}.

\subsection{Related Work}
Although user association problem has been extensively studied in the literature, the need for flexible, robust, and efficient solutions to handle the growing networks complexity is still an open research topic, especially in the context of dense mmWave networks. In fact, the user association is known to be a combinatorial and non-convex optimization problem, difficult to solve with the standard optimization tools. To deal with this challenge, Athanasiou \etal have designed a distributed algorithm to manage the user association \cite{athanasiou}. Their solution is sub-optimal as it ignores interference and does not consider the environment dynamics. Similarly, Lui \etal have formulated a non-cooperative game with local interactions for managing the beam pair selection between UEs  and BSs while targeting to maximize the network sum-rate \cite{lui}. However, this proposal requires information exchange among UEs, thus, inducing a large signaling overhead. Moreover, as in \cite{athanasiou}, the game also need to be replayed every time a relevant change occurs in the environment.

Recent advances on machine learning and reinforcement learning \cite{mao, busoniu2008} have enabled the design of more flexible algorithms for optimizing the user association. In this context, Zhou \etal have recently proposed a deep neural network (DNN) based user association scheme to maximize the network throughput subject to power and beam width constraints \cite{zhou}. Although with the use of DNN this solution reduces the computation overhead with respect to standard solutions, this approach is centralized and requires collecting the signal-to-noise ratio (SNR) information of all the links to construct the DNN inputs. This induces therefore a large amount of signaling overhead since the SNR values need to be gathered every time the environment changes. Moreover, the solution learned by the DNN is generated using a heuristic algorithm, which limits the system performance. 

Although training a deep neural network on mobile devices in a computation and energy efficient way is an ongoing research topic, notable efforts have already been made both in terms of hardware design and software accelerators (see \cite{deng2019deep, lee2019device} and references therein), which makes possible to move part of the optimization process at the user side. In this sense, Zhao \etal have tackled the problem of user association with a distributed multi-agent reinforcement learning (MARL) algorithm \cite{zhao}. Nevertheless, this work has not focused on mmWaves networks and has considered fully observable environment. Besides that, in contrast to our work, the solution proposed in \cite{zhao} lacks in scalability as the architecture of the proposed DNN depends on the total number of interacting UEs. In our previous works, we have demonstrated the feasibility of MARL in partially observable environment, by designing a scalable architecture that enables user association in static network (static channels) \cite{sana2019UA} and to optimize the handover cost in dense mmWave networks \cite{sana2019HO}. Now, we extend these previous studies to take into account more realistic assumptions, and let the user association optimally update with respect to the network dynamics, viz. fading, traffic, and interference.

\subsection{Main Contributions}
The contributions of this paper can be summarized as follows:
 \subsubsection{Sum-rate maximization in dense mmWave networks} We first formulate a user association problem to maximize the sum-rate of a mmWave network. In contrast to the existing works, we take into consideration both inter-cell and intra-cell interference and environment dynamics, which are characterized by the time-varying nature of the mmWave channels and the evolving data rate demand of UEs by using only local observations at each UE.
 
  \subsubsection{Multi-agent reinforcement learning based user association scheme} We cast the formulated user association problem into a multi-agent reinforcement learning task, where UEs, modeled as agents, collaborate to maximize the network sum-rate. In order to limit both signaling and computational complexity, the agents act as independent learners i.e., their decisions are independent of each other. We force UEs to act based only on partial observations and perceived rewards. This brings the benefit that a UE does not need to collect and process information related to other users. In this setting, we propose a deep recurrent Q-network (DRQN) architecture and the associated signaling protocol that enable UEs to learn an efficient network sum-rate association policy.
  
  \subsubsection{Policy distillation in small-scale dynamics} When the system dynamics change faster than the learning convergence time, online strategies cannot be successfully implemented. In particular, this is the case when the dynamic of the UE traffic requests changes in time and that the user association must be updated accordingly to avoid performance losses. To deal with these challenges, we design an offline \textit{distillation procedure} consisting in integrating experiences related to different scenarios in a single one, so that the users are able to adjust their association policy to abrupt changes of the environment.

The remainder of the paper is organized as follows. Section \ref{sysmodel-pbform} introduces the system model and formulates the optimization problem to be solved. Section \ref{MARL-scheme} details the proposed distributed algorithm and the associated signaling protocol. Section \ref{numerical-simu}  outlines the simulation results while Section \ref{conclusion} concludes the paper and gives some perspectives on future work.

\section{System Model and Problem Formulation}\label{sysmodel-pbform}

\subsection{System model}\label{systemmodel}
\begin{figure}[t]
    \centering
   \includegraphics[scale=0.9]{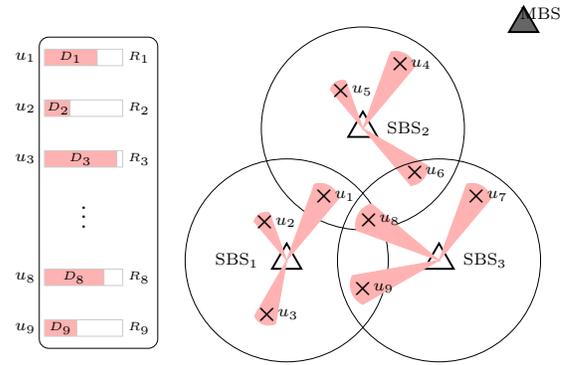}   
    \caption{A downlink network with $N_s=3$ SBSs, one MBS, and $K=9$ UEs. $\mathcal{U}_1=\{1,2,3,8,9\}$, $\mathcal{U}_3=\{6, 7, 8, 9\}$. As an example, $\mathcal{A}_1=\{1,2\}$, $\mathcal{A}_5=\{2\}$. Here, $D_j$ denotes UE $j$ data rate request and $R_j$ is the achievable data rate w.r.t. the selected BS.}
    \label{fig:arch}
\end{figure}

We consider a downlink network consisting of $N_s$ mmWave small cells and one macro cell jointly providing services to $K$ UEs as shown in \fig{fig:arch}. We denote by $\mathcal{A} = \{0, 1, 2, \dots, N_s\}$ the set of $N_s+1$ BSs in the network where $0$ indexes the macro base station (MBS),  which uses sub-6 GHz technology to enable ubiquitous network coverage. Also, we use $\mathcal{U}_i$ to indicate the set of UEs under coverage of the $i$-th BS; hence, $\mathcal{U} = \bigcup\limits_{i=0}^{N_s} \mathcal{U}_i=\{1, 2, \dots, K\}$ represents the set of all UEs in the network.

In this architecture with multi-radio access technologies, a UE may receive control signals from multiple BSs. Therefore, we define $\mathcal{A}_j = \{i,\quad d_{i,j} \leq \phi_i/2,~i\in\mathcal{A}\}$\footnote{$\mathcal{A}_j$ can also be derived based on links quality, e.g., the received signal strength indicator between UE $j$ and BS $i$ ($\mathrm{RSSI}_{i,j}$) should be greater than a predefined threshold $\kappa_j$, i.e., $\mathcal{A}_j = \{i,\; \mathrm{RSSI}_{i,j} \geq \kappa_j\}$.} as the set of BSs the UE $j$ could connect to, where $\phi_i/2$ is the cell radius of the BS $i$ and $d_{i,j}$ is the distance between BS $i$ and UE $j$. Note that $\mathcal{A}_j\neq\text{\O}, \forall j$ as a UE can always be associated with the MBS. Let $x_{i,j} \in \{0, 1\}$ be the binary association variable such that $x_{i,j} = 1$ when UE $j$ is served by the BS $i$ and $x_{i,j} = 0$ otherwise. Here we assume that each UE can only received data from one BS at a time. Moreover, we consider that each mmWave small cell BS (SBS) cannot serve more than $N_{i}$ UEs simultaneously, where $N_{i}$ is the maximum number of beams available at the SBS $i$. In our system model, we consider that the SBSs allocate all the available mmWave band to each served UE; in contrast, the MBS equally shares its band across the served UEs. Finally, we consider that the SBSs  and the UEs have already performed beam training and alignment mechanisms in advance and therefore are able to configure the appropriate beams when a data connection is set up. For instance, an initial access protocol based on the signal-to-interference-plus-noise ratio (SINR) can be used to complete this task \cite{Li2017}.
%%% system model %%%%

The downlink SINR between BS $i$ and UE $j$ is evaluated as follows:
\begin{equation}\label{eq:SINR}
    \mathrm{SINR}_{i,j}{}= \frac{\alpha_{i,j} P_{i}^{\mathrm{Tx}} {}G_{i,j}^{\mathrm{Tx}} {} G_{i,j}^{\mathrm{ch.}}{}G_{i,j}^{\mathrm{Rx}}}{I_{i,j}^{\mathrm{intra}} + I_{i,j}^{\mathrm{inter}} + N_0B_{i,j}}.
\end{equation}
Here, $P_{i}^{\mathrm{Tx}}$ is the transmit power of BS $i$, $N_0$ is the Gaussian noise power spectrum density, and $B_{i,j}$ the bandwidth allocated from the BS $i$ to the UE $j$. $ I_{i,j}^{\mathrm{inter}}$ and $ I_{i,j}^{\mathrm{intra}}$ are the inter-cell and the intra-cell interference on the link $(i,j)$, respectively. Since, in our system model, we consider only one MBS, which shares its bandwidth across the served UEs, $I_{i,j}^{\mathrm{intra}}=I_{i,j}^{\mathrm{inter}}=0$ for MBS-UE links. In contrast, for mmWave links, the inter-cell and intra-cell interference are due to the overlapping of different beams serving distinct UEs, as we do not specifically optimize the beamformers. Their expressions  are given as follow:
\begin{IEEEeqnarray}{ll}\label{eq:Interf}
     I_{i,j}^{\mathrm{intra}}{}&= \sum_{j'\in \mathcal{U}_i/\{j\}}  x_{i,j'} \alpha_{i,j}  P_{i}^{\mathrm{Tx}} G_{i,j'\rightarrow j}^{\mathrm{Tx}} G_{i,j}^{\mathrm{ch.}} G_{i,j\leftarrow i}^{\mathrm{Rx}},\IEEEyesnumber\IEEEyessubnumber\\
     I_{i,j}^{\mathrm{inter}}{}&=\sum_{(i',j')\in \mathcal{V}}  x_{i',j'} \alpha_{i',j}  P_{i'}^{\mathrm{Tx}} G_{i',j'\rightarrow j}^{\mathrm{Tx}} G_{i',j}^{\mathrm{ch.}} G_{i,j\leftarrow i'}^{\mathrm{Rx}},~~\IEEEyessubnumber
\end{IEEEeqnarray}
 where $\mathcal{V}=\mathcal{A} \backslash\{i, 0\}\times\mathcal{U}\backslash\{j\}$ is the set of inter-cell interfering links with respect to link $(i,j)$. $\alpha_{i,j}$ represents the small scale fading coefficient, which in case of mmWave channels is typically modelled with a $m$-Nakagami\footnote{For sub-6 GHz channels, we consider $m=1$, which is equivalent to Rayleigh fading.} distribution \cite{chevillon2018}.

Here, $G_{i,j}^{\mathrm{ch.}}$ denotes the channel gain between BS $i$ and UE $j$. Let $s$ denote the type of BS, i.e., $s\in\{\mathrm{SBS}, \mathrm{MBS}\}$. Then, we adopt a distance-based channel model \cite{Rappaport2017}, which captures the effect of path loss and shadowing as follows:
 \begin{equation}
 G_{i,j}^{\mathrm{ch.}}(\mathrm{dB}){}={} 20 \mathrm{log}_{10}\left(\frac{4\pi d_{0,s}}{\lambda_s}\right) + 10\eta_s \mathrm{log}_{10}\left(\frac{d_{i,j}}{d_{0,s}}\right) + X_{i,j}^{\sigma_s}, \label{eq:ploss}
 \end{equation}
 where $d_{0,s}$ is the close-in-free-space reference distance, $\eta_s$ is the path loss coefficient, $\lambda_s$ the wavelength, and $X_{i,j}^{\sigma_s}$ the log-normal shadowing with a variance equal to $\sigma_s^2$. In particular, $d_{0,s}$, $\eta_s$, $\lambda_s$ and $\sigma_s^2$ depend on the radio technology; $G_{i,j}^{\mathrm{Tx}}$ and $G_{i,j}^{\mathrm{Rx}}$ are the transmitter and receiver antenna gain in the communication link, respectively. 
We denote with  $G_{i',j'\rightarrow j}^{\mathrm{Tx}}$ the transmitter gain of the interfering link $(i',j')$ on UE $j$ related to the communication between BS $i'$ and UE $j'$; its value depends on the orientation of the beam used by BS $i'$ for transmitting to UE $j'$ and the position of the UE $j$.
Moreover, $G_{i,j\leftarrow i'}^{\mathrm{Rx}}$ indicates the receiver gain of the  UE $j$, communicating with BS $i$, with respect to the interring BS $i'$; its value depends on the orientation of the beam used by UE $j$ to receive from BS $i$ and the position of BS $i'$.

In addition, we assume that the backhaul network has sufficient capacity so that we neglect its impact when optimizing the user association. However, this framework can be easily extended for the scenarios where a backhaul with limited capacity can degrade the users' performance, e.g., following the approaches in \cite{Domenico2013BackHaul, Sapountzis2017} or by limiting the number of UEs served by each SBS.
From the above definitions, the achievable communication rate between BS $i$ and UE $j$ is given by the Shannon capacity:
\begin{equation}\label{SINR}
 R_{i,j}(t)=B_{i,j} \mathrm{log}_2\left(1 + \mathrm{SINR}_{i,j}(t)\right).
\end{equation}
 
 Moreover, in our model, we take into account the UE traffic when computing the network throughput. Accordingly, we define $D_j(t)$ as the \textit{data rate demand} of UE $j$ at time step $t$, which follows a Poisson distribution with intensity $\overline{D}_j =  \mathbb{E}\left[D_j(t)\right]$.

\subsection{Problem formulation}
Given the UE $j$ with a traffic demand $D_j(t)$, the effective data rate exchanged with BS $i$ at the time $t$ is $\mathrm{min}\Big(D_j(t), R_{i,j}(t)\Big)$. Let $R(t)$ be the total network sum-rate, which is defined as follows:
\begin{align}\label{def-rate}
R(t)=\sum_{i\in \mathcal{A}}\sum_{j \in \mathcal{U}}x_{i,j} \mathrm{min}\Big(D_j(t), R_{i,j}(t)\Big).
\end{align}
 We formulate the user association problem to maximize the network sum-rate \eqref{def-rate} as follows:
\begin{IEEEeqnarray}{lCl}
&\underset{\{x_{i,j}\}}{\mathrm{maximize}}~~& R(t), \IEEEyesnumber\IEEEyessubnumber\label{eq:Obj}\\
&\mathrm{subject~to~}& x_{i,j} \in \{0,1\},\IEEEyessubnumber\label{eq:C1}\\
	 &{}&\sum_{j \in \mathcal{U}_i}x_{i,j} \leq N_{i}, ~ i \in \mathcal{A}\backslash \{0\},\IEEEyessubnumber\label{eq:C2}\\
	 &{}&\sum_{i \in \mathcal{A}_j}x_{i,j} = 1, ~ j \in \mathcal{U}.\IEEEyessubnumber\label{eq:C3}
\end{IEEEeqnarray}
The constraint (\ref{eq:C1}) ensures that the decision variables $x_{i,j}$ are binary. The constraint (\ref{eq:C2}) highlights that a given SBS $i$ can use at most $N_{i}$ beams at the same time. In our architecture, we assume that the MBS can simultaneously support any number of requests by equally sharing the available band across the served UEs. Finally, constraint (\ref{eq:C3}) indicates that, in our setting, each UE is associated with exactly one BS.

One has to notice that the objective function (5a-5d) is non-convex. Indeed, the association of a given UE $j$ with a given BS $i$ depends on its $\mathrm{SINR}_{i,j}$ value. However, by observing \eqref{eq:SINR} the expression of the SINR also depends on the association of other users through the interference terms in the denominator. These cross-dependencies combined with the binary decision variables make the optimization problem non-convex and NP-hard, hence, difficult to solve with conventional optimization frameworks \cite{shen2018}. The difficulty is exacerbated when considering the UEs traffic as it introduces a non-linearity through the $min(.,.)$ function. In the next sections, we propose to tackle the aforementioned issues with deep reinforcement learning (DRL). 

\section{Proposed Solution via Deep multi-agent Reinforcement Learning}\label{MARL-scheme}

In this section, we describe the proposed solution to solve the user association problem in a dynamic environment. First, to limit the complexity of the proposed solution, we cast this problem to a multi-agent DRL framework, where each user independently learns the optimal policy. This solution is distributed based on multi-agent reinforcement learning. Then, to characterize the associated overhead, we show the signaling messages required to implement the proposed solution in a practical system.

\subsection{Background on multi-agent reinforcement learning}
The proposed user association framework is based upon the cooperative MARL approach, but in our solution, we do not exploit inter-agent communications. Cooperative MARL refers to framework in which agents learn to coordinate together to achieve a common objective \cite{bucsoniu2010multi}. In such a framework, agents learn by interacting with a shared environment following a Markov Decision Process (MDP). Basically, an MDP is defined as tuple $(\mathcal{S}, \mathcal{A}, \mathcal{T}, \mathcal{R})$, in which $\mathcal{S}$ denotes the state space, $\mathcal{A}$ is the action space, $\mathcal{T}(s,a,s')$ the probability of transitioning from state $s$ to state $s'$ after taking action $a$, which results in an immediate reward $\mathcal{R}(s,a,s')$. The problem for agents in MDP is to find the optimal policy $\pi^{*}:\mathcal{S}\rightarrow\mathcal{A}$ that maximizes the expected sum of the perceived rewards. To solve this problem, Q-Learning is a widely used model-free algorithm that estimates the Q-values $Q(s,a)$, which are the expected maximum sum of rewards perceived at state $s$ when taking action $a$.

Recent exploits in deep learning enable to approximate the Q-function in order to deal with complex problems with deep Q-network (DQN): $Q(s,a) \approx Q(s,a; \theta)$, where $\theta$ is the set of neural network parameters \cite{mnih2015humanlevel}. DQN relies on experience replay to speed up and stabilize the training process \cite{mnih2015humanlevel}. At each time $t$, from a state $s(t)$, agent takes an action $a(t)$ following a policy (e.g., $\epsilon$-greedy) that brings it to state $s(t+1)$ with an immediate reward $r(t)$. The resulting experience $e(t) = \{s(t), a(t), r(t), s(t+1)\}$ is stored into an experience replay memory $\mathcal{M}$ from which a mini batch of experience $\mathcal{B}$ is sampled every iteration during the learning phase. In this phase, the weights of the DQN are iteratively updated using stochastic gradient descent (SGD) on mini batches in order to minimize the loss function:
\begin{equation}
\label{lossfunc}
\mathcal{L}(\theta){}={} \mathbb{E}_{e(t)\sim\mathcal{B}}\left[\delta(t)^2\right].
\end{equation}
In \eqref{lossfunc}, $\delta(t) = y(t) - Q(s(t), a(t);\theta)$ denotes the temporal difference (TD) error where the $\gamma$-discounted target value is computed as follows:
\begin{equation}\label{target-value}
y(t) = r(t) + \gamma~\underset{a'}{\mathrm{max}}~Q(s(t+1),  a'; \theta).
\end{equation}
Finally,  knowing the optimal parameters $\theta^*$, the optimal policy is given by:
\begin{align*}
 \begin{array}{ll}
     \pi^{*}: & \mathcal{S}\rightarrow\mathcal{A}  \\
     {}& s\rightarrow \underset{a\in\mathcal{A}}{\argmax}~Q(s,a;\theta^*).
\end{array}
\end{align*}
In general, there can be some states where the outcome is the same regardless of the action the agent could take; therefore, it is not always necessary to determine the state action value at a given state $s$, $Q(s, a; \theta)$, for every action. For instance, when playing a video game consisting in moving left or right to avoid objects, trying to decide whether the optimal action is to move left or right is totally useless if there is no threatening object in sight. Another example is when a UE is located at the same distance from two BSs that can provide it with the same throughput. In that case, there is not a single optimal action as the result will be the same whatever BS is selected Based on this intuition, Wang \etal have introduced the notion of dueling network where $Q(s, a;\theta)$ is decomposed into a state value $V(s;\theta) =  \mathbb{E}[Q(s,a; \theta)]$ and the \textit{advantage} of the corresponding action $A(s,a;\theta)$  \cite{wang2015dueling}. That is, 
\begin{equation}
Q(s, a;\theta) = V(s;\theta) + A(s,a;\theta).
\end{equation}
The first term is action-less and is inherent to the state while the second measures the goodness of the action in that state. Dueling network shows that learning the DQN by estimating separately the state value and the advantage values can enable notable improvement in the agent policy.
 
In addition, it is important to highlight that, in distributed deep MARL, each agent maintains its own DQN, i.e., its own policy, while sharing its environment with other agents. Typically, in this context, either each agent acts in a selfish way, learning a policy that optimizes its own performance, or aims to determine a global optimal policy, which maximizes the system performance.  One major issue that arises with MARL is the problem of non-stationarity due to multiple agents interacting simultaneously with the environment. As a result, this can lead to \textit{shadowed equilibria} where an agent's locally optimal action could end up being globally sub-optimal \cite{matignon2012independent}. In the following, we focus on cooperative MARL, meaning that agents also share a common joint reward and propose a solution to deal with this \textit{shadowed equilibria}.

\subsection{Proposed framework}
\begin{figure}
\centering
\includegraphics[scale=0.49]{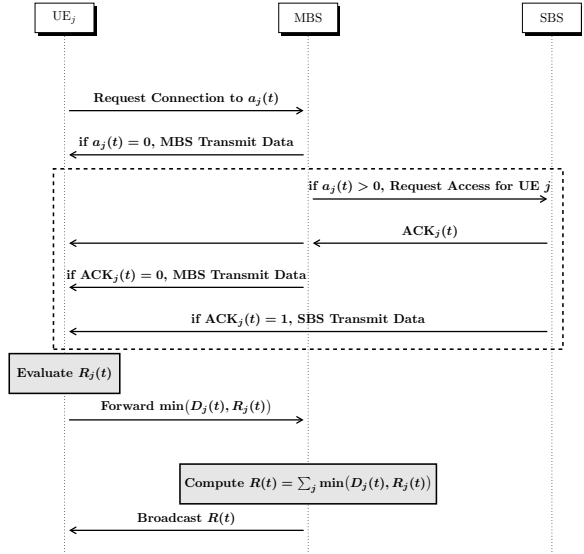}
\caption{Message sequence chart of the proposed mechanism.}
\label{fig:infoex}
\end{figure}

In this section, we define the proposed MARL framework to solve the optimization problem \eqref{eq:Obj}-\eqref{eq:C3}. In this framework, we model UEs as agents and assign them a common objective, i.e., maximizing the network throughput. In our setting, a UE, based on its local observations, selects and requests service from a target BS, which accepts or rejects the connection request by sending an acknowledgment (ACK) signal depending on the available resources. 

As described in \fig{fig:infoex}, each UE $j$ starts by identifying the set of BSs $\mathcal{A}_j$ it may connect to. Note that, in practical systems, the size of this set is limited to reduce the complexity of mobile devices. $\mathcal{A}_j$ also defines the UE action space, meaning that the action $a_j(t) \in \mathcal{A}_j$ denotes the index of the BS to which the UE $j$ requests a connection at time $t$. Then, in every time step, each UE $j$ takes an action $a_j(t)$ and informs the MBS of its choice. If the UE is requesting a connection from the MBS, i.e., $a_j(t)=0$, the request is automatically granted\footnote{Recall that we assume that the MBS is able to simultaneously serve all the active UEs by equally sharing its band across them.} and the communication is set up. Otherwise, the MBS forwards the connection request to the corresponding SBS. Depending on the overall received requests and constraint \eqref{eq:C3}, the SBS notifies both the UE and the MBS with an $\mathrm{ACK}_j(t)$ signal. If $\mathrm{ACK}_j(t)=1$, the SBS grants a connection to the UE; otherwise, the MBS establishes the default data link with the UE $j$. Next, each UE $j$ evaluates the perceived data rate, i.e., $\mathrm{min}\left(D_j(t), R_{a_j(t),j}(t)\right)$ and forwards this value to the MBS. Then, the MBS computes the network sum-rate $R(t)$. Finally, the MBS broadcasts the total throughput to each UE, which uses it to evaluate the goodness of its policy $\pi_j(t)$ and to update it accordingly.

Following this process, we define the history $\mathcal{H}_j(t)$ of UE $j$ as the set of all actions, observations, and measurements collected up to time $t$ \cite{naparstek} :
\setlength{\arraycolsep}{0.0em}
\begin{eqnarray}
    \mathcal{H}_j(t)&{}={}&\Big\{a_j(\tau),~\mathrm{ACK}_j(\tau),~\mathrm{RSSI}_{a_j(t),j}(\tau),\nonumber\\
&&~~D_j(\tau),~R_{a_j(t),j}(\tau),~R(\tau)\Big\}_{\tau=1}^t.
\end{eqnarray}
Hence, the policy of UE $j$ at time $t$, $\pi_j(t)$, is a mapping from its history $\mathcal{H}_j(t-1)$ to a probability mass function over its action space $\mathcal{A}_j$. Therefore, each UE takes its actions following its own strategy without being aware of the actions taken by the other UEs.

A key feature of the proposed approach is that in contrast to MDPs, here, the decision of the $j$-th UE is based only on its \textit{local} state observation $o_j(t)=\big\{a_j(t-1), R_{a_j(t-1),j}(t-1), R(t-1), {\mathrm{ACK}_j(t-1), \mathrm{RSSI}_{a_j(t-1),j}(t), D_j(t)} \big\}$. It is worth to note that $o_j(t)$ carries information related to the previous action/reward, already available at the UE side, and new local information (the RSSI and the data demand $D_j(t)$). Specifically, each UE makes association decisions based on how well its previous actions performed. The only observation that implicitly coordinates the actions of the multiple UEs is the network sum rate, which serves as a signal to each UE as to whether their local actions are beneficial to the overall network objective. Please note that the overall network objective may increase or decrease due to the actions of multiple UEs, thus it is not a perfect signal in the sense that it does not tell each UE exactly the consequence of its own specific action. Yet, our goal is that, using DRL, each UE is able to learn over time its optimal policy.

It is noteworthy that the size of the state observation of a given UE does not scale with the number of UEs in contrast to other works in the literature, as \cite{zhao}. This allows us to build general DQNs that can be used in different network scenarios; that is, if a UE leaves or joins the network, there is no need to change the DQN architecture.  Moreover, $o_j(t)$ is a partial observation of the true state $s(t)$, which includes all the observations of other agents. In the literature, the optimization in partially observable environments is addressed as a multi-agent partially observable Markov decision process \cite{omidshafiei2017}. Partial observability, in addition to non-stationarity issues, make MARL an even more complex task. To tackle this problem, Omidshafiei \etal successfully applied hysteretic Q-learning (first introduced by Matignon \etal\cite{matignon2007hysteretic}) with partial observability \cite{omidshafiei2017}. They empowered the DQNs with recurrent neural networks (RNN) to obtain deep recurrent Q-networks (DRQNs), which serves as a basis to our proposed algorithm.

\begin{figure*}[!t]
    \centering
    \includegraphics[scale=0.8]{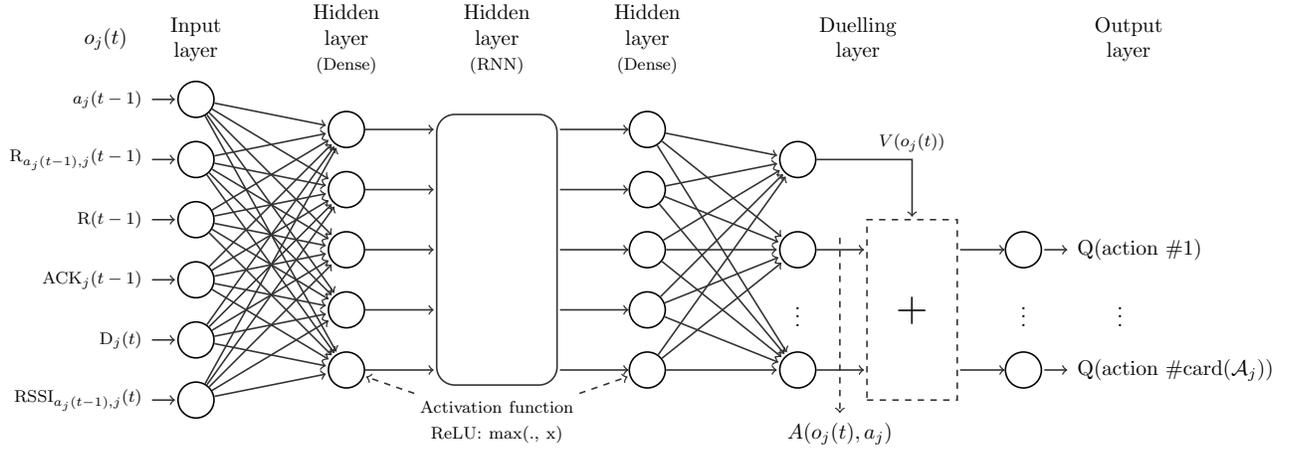}
    \caption{Illustration of the architecture of the proposed DRQN.}
    \label{fig:qnet}
\end{figure*}

\subsection{Hysteretic Deep Recurrent Q-Network}

In the hysteretic deep recurrent Q-network (HDRQN) algorithm\footnote{In the following, we use HDRQN when we refer to the proposed algorithm and DRQN to the related neural network architecture.}, each UE $j$ acts as an \textit{independent learner} and maintains its own DRQN $Q_j(o_j(t), h_j(t-1), a_j(t); \theta_j)$. \fig{fig:qnet} describes the proposed DRQN, which is composed by one input layer, two fully connected hidden layers, one RNN hidden layer, a dueling layer, and an output layer. The UE's local state information $o_j(t)$ and the estimated state action value $Q_j(\cdot;\cdot)$ define respectively the input layer and the output layer of the DRQN (Section \ref{numerical-simu} provides more details on the proposed DQRN). We use $h_j(t-1)$ to represent the internal state of the RNN hidden layer and $\theta_j$ to define the UE's local DRQN weights. The use of RNNs allows to aggregate past information (previous observed states, i.e., the history $\mathcal{H}_j(t)$) in agent decision making process, which is shown to improve the average reward perceived when dealing with partial observability \cite{hausknecht2015deep}. Indeed, in partially observable environment, each agent makes its decision relying on the observation $o_j(t)$ instead of the true state $s_j(t)$, which is unknown. From $o_j(t)$ solely, the agent  may have a partial perspective of the environment. In this case, the commonly used Vanilla DQN may not be effective \cite{hausknecht2015deep}, specifically in multi-agent {scenarios}, where each agent is unaware of the behavior of its teammate. Hence, {we extend} the {baseline} Vanilla DQN with RNN to infer the underlying state $s_j(t)$ from agent past observations, i.e., its history $\mathcal{H}_j(t)$ \cite{hausknecht2015deep}.

The experience of the $j$-th UE ${e_j(t) = \left\{o_j(t), a_j(t), r_j(t), o_j(t+1)\right\}}$ is stored into a local memory buffer $\mathcal{M}_j$. In order to further stabilize the learning process, synchronized sampling strategy (called \textit{concurrent experience replay trajectories} (CERTs)) is adopted \cite{omidshafiei2017}. In other words, during the training, mini batches of experiences of the same time steps are sampled across agents to update the local DRQN weights in order to minimize the hysteretic loss function:
\begin{equation}
\label{lossfunction}
\mathcal{L}_j(\theta_j){}={} \mathbb{E}_{e_j^b(t)\sim\mathcal{B}_j}\left[\left(w_j^b\delta_j^b(t)\right)^2\right].
\end{equation}
In \eqref{lossfunction}, $b$ indexes an entry in the mini batch of experiences $\mathcal{B}_j$, $\delta_j^b(t) = y_j^b(t) - Q_j(o_j^b(t), h_j^b(t-1), a_j^b(t); \theta_j)$ is  the TD error with respect to the target value 
\begin{equation}
\label{target}
y_j^b(t) = r_j^b(t) + \gamma~\underset{a'}{\mathrm{max}}~Q_j(o_j^b(t+1), h_j^b(t), a'; \hat{\theta}_j).
\end{equation}
Here, $\hat{\theta}_j$ represent the weights of the target DRQN, which is updated less frequently to improve learning stability \cite{mnih2015humanlevel}. 

In MARL, the agents' reward is the result of their joint actions. Accordingly, an agent experience $e_j(t) $ is \textit{positive}, if the associated TD error $\delta_j^b(t)$ in \eqref{lossfunction} is positive,  i.e., the perceived global reward is better than the previous rewards independently of the optimality of the agent local action. This does not necessarily imply that the agent's strategy is converging toward the optimal solution but the network performance is improving over time. In contrast, a \textit{negative} experience results in an agent receiving a lower reward after taking an action that was fruitful in the past. This can be caused by the agent's action being non-optimal or more likely by the others agents' behavior. That is, an agent that has taken a local optimal action, may receive a lower reward because of the bad choices of other agents. Therefore, negative experiences can be very detrimental in MARL as they may mislead the agent to change its optimal strategy. Consequently, an agent may stabilize its strategy by paying less attention to negative experiences.

This is the idea introduced by hysteretic Q-learning: the neural network weights are updated via SGD with two distinct learning rates $\alpha\mu$ and $\beta\mu$ ($\beta \ll \alpha \leq 1$), where $\mu$ is a based learning rate and $\alpha$ and $\beta$ are control factors. When the TD error is positive, the learning rate $\alpha\mu$ is used; otherwise, $\beta\mu$ is considered. This leads to optimistic updates that give more importance to positive experiences \cite{matignon2007hysteretic}. To implement the hysteretic learning in conventional machine learning libraries, we set $\mu$ as the fixed learning rate and scale the TD-error $\delta_j^b(t)$ in \eqref{lossfunction} as follow:
\begin{equation}
\label{hystereticweights}
w_j^b{}={} \left \{
				\begin{array}{r l}
					\alpha, &\quad \text{if } \delta_j^b(t) \geq 0\\
					\beta, &\quad \text{otherwise}
				\end{array}.
			\right.
\end{equation}

\subsection{Definition of the reward function}

The maximum value of the network sum-rate, and hence, the optimal user association is unknown to the agents at the beginning of the learning phase. In other words, there is no explicit or  predefined terminal state that agents are aware of and toward which they have to converge to. Accordingly, we treat this learning problem as a continuing task over a time horizon $T_e$. That is, the agents keep updating their policies as long as it improves the perceived reward.

\begin{definition}
We define the beam \textit{collision} as the event corresponding to a given SBS $i$ receiving more requests than the number of beams it can set up i.e., $N_i$.
\end{definition}

This may occur since all UEs are requesting connections simultaneously. However, in our proposed framework, we aim at effectively training agents in order to distribute the network load and properly leverage the advantages of network densification. Consequently, when a collision happens during the training phase, we punish all UEs by setting the instantaneous reward to zero\footnote{Note that setting the reward to zero as mentioned, is simply to discourage agents from colliding. However, in practice implementation of this framework, during execution time, one may choose between the colliding UEs, which UEs to serve. This selection can be made either randomly or based on RSSI.}. As a result, we define the reward function of UE $j$ in \eqref{target} as:
\begin{equation}\label{eq:reward-def}
r_j(t) = \left \{
				\begin{array}{r l}
					R(t), &\quad \text{if there is no collision}\\
					0, &\quad \text{otherwise}
				\end{array}.
			\right.
\end{equation}
During the learning, each UE $j$ builds its policy $\pi_j$ depending on its data rate requirement, the experienced SINR, the network sum-rate, and whether its requests cause a collision to maximize the accumulated discounted reward:
\begin{equation}
G_j(t) = \sum_{\tau=t+1}^{T_e} \gamma^{\tau-t-1}r_j(\tau),
\end{equation}
where the discounting factor $\gamma$ is such that $0\leq\gamma<1$. Taking $\gamma=0$ leads to myopic (instantaneous) network throughput maximization. In case of dynamic scenarios, it is better to consider $\gamma \neq 0$ to take into account the dynamic nature of the environment: there is no need to change the current user association at time step $t$ due to a low reward perceived because of the environment dynamics if at the next time step the system will recover its \textit{equilibrium}. This consideration also make sense in a practical system where changing the association too often can also induce excessive overhead.

As defined, the reward perceived by the agents continuously varies with the environment stochasticity viz. fading, shadowing, interference, traffic, and noise. Accordingly, this reward setting can lead to many optimal or quasi-optimal \textit{equilibria}, which is a major issue as it results in agents laboriously trying to converge \cite{matignon2012independent}. Algorithm \ref{algo:UEassos} presents the proposed training procedure to deal with these challenges. It this noteworthy that the parts of this algorithm highlighted in gray can be executed in parallel across all UEs.

%%%%%%%%%%%%%%%%%%%% UE Association algorithm %%%%%%%%%%%%%%%%%%%%%%%%%%%
%\resizebox{0.8}{
%\begin{minipage}{0.5\textwidth}
\begin{algorithm}
\footnotesize
\DontPrintSemicolon
\While{$t<T_e$}{
	\tikzmk{A}\For{$j \in \mathcal{U}$}{
	     Observe state $o_j(t)$. \;
	     With probability $\epsilon$ pick random action $a_j(t)$ in $\mathcal{A}_j$;  otherwise, $a_j(t) \gets \underset{a'\in\mathcal{A}_j}{\argmax}~Q_j(o_j(t), h_j(t-1), a'; \theta_j)$.\;	     
	    \eIf{{$a_j(t)\neq 0$ and connection granted}}{	       
	        $\mathrm{ACK}_j(t) \gets 1$. \tcc*[r]{{\footnotesize the UE is requesting a connection to a SBS}}
	    }{
	        $\mathrm{ACK}_j(t) \gets 0$. \;
	        Automatically redirect to the MBS. \;
	    }
	    Measure $R_{a_j(t),j}$. \;
	}\tikzmk{B}
	\boxit{gray}{A}{B}
	$R(t) \gets 0$ \;
	\For{$i \in \mathcal{A}$}{
	\eIf{$\sum_j \mathbbm{1}_{a_j(t)=i} > N_i$}{
	    $R(t) = 0$.  \tcc*[r]{{\footnotesize collision}}
	    Break. \;
	}{
	    $\displaystyle R(t) = R(t) + \sum_{j\in\mathcal{U}_i} \mathrm{1}_{a_j(t)=i}\mathrm{min}(R_{i,j}, D_j)$. \;
	}
	}
	\tikzmk{C}\For{$j \in \mathcal{U}$}{
	    Observe the new state $o_j(t+1)$. \;
	    Store experience $e_j(t)$ into $\mathcal{M}_j$. \;
	    Samples a batch of experiences from  $\mathcal{M}_j$.\;
	    Compute the target value $y_j^b(t)$. \;
	    Performs a gradient descent step on $\delta_j^b(t)$ with respect to $\theta_j$. \;
	    Periodically reset $\hat{\theta}_j \gets \theta_j$. \;
	}\tikzmk{D}
	\boxit{gray}{C}{D}
	$t = t + 1$.\;
}
\caption{User Association: Training Procedure}\label{algo:UEassos}
\end{algorithm}
%\end{minipage}

\subsection{Adaptability with respect to different service requests}
\label{section:distillation}

\begin{figure}[t]
    \centering
    \includegraphics[scale=0.65]{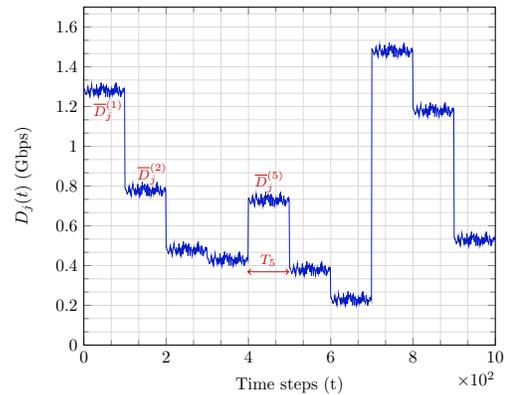}
    \caption{Example of the variation of UE $j$ service request with time.}
    \label{fig:traffic-evolv}
\end{figure}
We now focus on a more realistic scenario, where the service requests of the UEs can change over time, e.g., from video streaming to virtual reality applications. We model this change by abruptly modifying the intensity of the Poisson distribution that characterizes the UE traffic, i.e., for UE $j$, $\overline{D}_j(t)$ is now time-dependent (see \fig{fig:traffic-evolv}). This increases the non-stationarity of our system and makes the learning process more challenging. To deal with this, the agents may keep updating their policies online, to adapt them to an eventual drastic change in the environment dynamics. This approach may lead to good performance if the convergence time of the algorithm is sufficiently shorter than the time during which the system is stationary. However, in a multi-agent system this condition is unlikely satisfied and thus, we design an offline training strategy that allows the agents to perform well during the execution time even in strong non-stationary conditions. 

Let us assume that the time horizon $T_e$ can be divided into $P$ time intervals $T_p$ such that $\sum_{p=1}^P T_p=T_e$, where the intensities  $\overline{D}_j(t),~\forall j\in\mathcal{U}$ remain constant. Accordingly, we denote by $\overline{D}_j^{(p)}$ the average data rate requested by UE $j$ in the time interval $p$. Then, we define a task $\mathcal{T}_p$ as the set of the UEs' traffic requests during the time interval $p$:
\begin{equation}
    \mathcal{T}_p=\left\{\overline{D}_1^{(p)}, \overline{D}_2^{(p)}, ..., \overline{D}_K^{(p)}\right\}.
\end{equation}
In our setting, each agent does not have the global knowledge of each task specifications;  in fact, a UE is unaware of the data rate demands of the other UEs. However, we aim to derive, for each user, a unique policy that performs well in any task. 
This problem falls in the context of  the so-called multi-task reinforcement learning (MTRL) \cite{rusu2015policy}, where policy distillation consolidates multiple task-specific policies into a single policy (see Algorithm \ref{algo:distillation}). 

\begin{algorithm}[!t]
\footnotesize
\DontPrintSemicolon
\For{$p=1, \dots, P$}{
    Initialize $o_j = \{0\}$. \;
    Select policy $\pi_j(\mathcal{T}_p)$. \;
    \For{$t=0, \dots, T_p$}{
        Observe new state $o_j(t)$. \;
        Using  the expert policy $\pi_j(\mathcal{T}_p)$ takes $a_j(t)$. \;
        Get $Q_j(o_j;\theta_j)$. \;
        Store $\langle o_j(t), Q_j(o_j(t);\theta_j)\rangle$ into a memory $\mathcal{M}_j$. \;
    }
}
Initialize the distilled DRQN weights $\theta_j^D$. \;
Perform supervised learning using $\mathcal{M}_j$. \;
\caption{Distillation Procedure for UE $j$}\label{algo:distillation}
\end{algorithm}

Specifically, with this mechanism, for every task, we run Algorithm \ref{algo:UEassos} to collect the agents task-specific policies $\pi(\mathcal{T}_p)$; that is, we derive as many policies as there are tasks for any single agent. Then, for every agent $j$ and task $p$, we execute the related policy for a time $T_p$ and we store all the collected observations/action values $\langle o_j(t), Q_j(o_j(t);\theta_j)\rangle_p$ into a memory $\mathcal{M}_j$. Later, for each UE $j$, we conduct a supervised learning on the generated database $\mathcal{M}_j$ to learn a distilled policy $\pi_j^D$ trough a single DRQN (having the same architecture as in \fig{fig:qnet} with parameters $\theta_j^D$) trained via a tempered Kullback-Leibler (KL) divergence loss function:
\begin{equation}
    \mathcal{L}(\theta_j^D) = \mathbb{E}_{\mathcal{M}_j} \left[\mathrm{softmax}\left(\frac{Q_j}{\tau}\right)\mathrm{log}\left(\frac{\mathrm{softmax}\left(\frac{Q_j}{\tau}\right)}{\mathrm{softmax}\left(Q_j^D\right)}\right)\right],
\end{equation}
where the temperature $\tau$ controls the way the knowledge is transferred from the expert policies to the distilled policy \cite{rusu2015policy}. Increasing the temperature softens the Q-values which may prevent the distilled agent to take the same actions as the expert. In contrast, when the temperature get decreased, Q-values becomes more and more sharpened ensuring more knowledge distillation. Therefore, $\tau$ is typically set as a small positive value \cite{rusu2015policy}. 

\section{Numerical results}\label{numerical-simu}
\begin{figure}[!t]
    %----------------------     DIAGRAMS     -------------------------%
    \centering
     \includegraphics[scale=0.7]{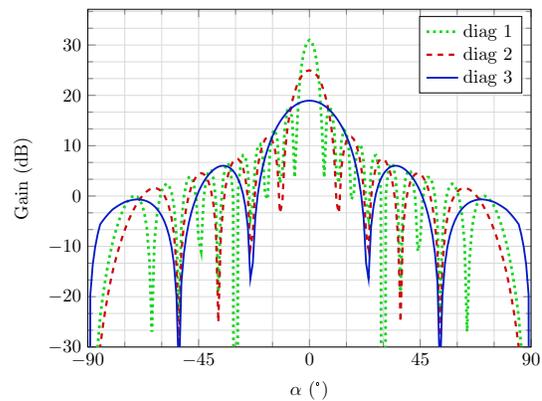}
    \caption{Simulated TX/RX antenna gain radiation pattern for an array of $20\times20$ (diag 1), $10\times10$ (diag 2), $5\times5$ (diag 3) elements operating at $28$ GHz \cite{mailloux2017phased}.}
    \label{fig:diagrams}
\end{figure}
In this section, we demonstrate the effectiveness of the proposed HDRQN-based user association by comparing its performance with two centralized benchmarks of the literature:

   \textbf{Max-SNR:} Each UE is associated with the BS providing the maximum SNR taking into account the constraint on the number of beams per BS (see \eqref{eq:C3}). Since this method does not consider interference, it has limited performance especially in case of a dense networks.
    %\item 
   
    \textbf{Heuristic:} Proposed in \cite{zhou}, this algorithm starts by ordering all the possible associations according to their respective SNR values, which do not consider interference. Then, following this order, the algorithm goes from one potential association to the following one and validates it if it increases the network sum-rate $R(t)$ in \eqref{def-rate}.  Although the evaluation of $R(t)$ takes the interference and UEs traffic into account, the output of the algorithm is mainly based on the SNRs ordering, which may prevent to reach a global optimum. This approach is recalled in Algorithm \ref{alg:heuristic} with minor modifications compared to the original one since power and beam width constraints are not considered in this study.
\begin{algorithm}
\footnotesize
\DontPrintSemicolon
Set $x_{i,j} = 0, ~\forall j\in\mathcal{U}, i\in\mathcal{A}_j$. \;
Get the $SNR_{i,j}$ and sort it in descending order into $\mathcal{Z}=\{z_1, z_2, ..., z_p, ..., z_P\}$ with ${P=\sum_j\card{\mathcal{A}_j}}$. Let $\delta$ be the transformation (defined by the sort) such that $\delta(i,j)=p$. That is, $z_p = z_{\delta(i,j)}\leftarrow SNR_{i,j}$.  \;
Set $R^{1}(t)=0$.  \;
\While{$p\leq P$}{
    Set $x_p=1$.  \;
    Compute $R^p(t)$. \tcc*[r]{{\footnotesize $R^p(t)$ is the sum-rate at iteration $p$.}}
    \eIf{$R^p(t) > R^{p-1}(t)$ and \textit{a beam is available}}{        
        Let $x_p$ unchanged. \tcc*[r]{{\footnotesize means that activating this link improves the sum-rate.}}
    }{
        Reset $x_p=0$. \;
    }
}
Apply $\delta^{-1}$ to recover which links (i,j) are active. \;
\caption{Heuristic scheme: Centralized User Association}\label{alg:heuristic}
\end{algorithm}

In the following, we start by analyzing the complexity of the proposed method with respect to the two baselines. Then, we study the effect of the hysteretic parameter on both convergence speed and achievable sum-rate. Also, we evaluate the effectiveness of collision cost in limiting collision events and improving network sum-rate. We continue assessing the performance of our scheme in both static and dynamic scenarios. Finally, we conclude the evaluation by demonstrating the adaptive property of the proposed algorithm.

We consider that UEs and SBSs communicate in the mmWave band at a carrier frequency of 28 GHz using the same phase-array antenna. To evaluate three different interference scenarios, we consider distinct antenna gain radiation patterns (see \fig{fig:diagrams}), which correspond to distinct number of antenna elements in the phase array. Larger the array, thinner the beam\footnote{Note that increasing the number of antenna elements also increases the antenna's size, which further increases the hardware complexity.}. In all tests, three small cells are deployed inside the macro cell. UE and small cell locations follow the 3GPP recommendations \cite{TR36872}. Table \ref{simu-params} summarizes the network parameters.
\begin{table}[!t]
    %\centering
    %\begin{minipage}{.5\linewidth}
    \centering
    \caption{Simulations parameters.}
    \label{simu-params}
    \scalebox{0.9}{
    \renewcommand{\arraystretch}{1.3}
    \begin{threeparttable}
    \begin{tabular}{l||c||c}
       \hline
        & Macro cell \protect\cite{TR36872} & Small cell \protect\cite{Rappaport2017}\\
       \hline
       \hline
        Parameters & \multicolumn{2}{c}{Values} \\
       \hline
        Carrier frequency, $f_s$ & 2.0 GHz & 28 GHz\\
        \hline
        Bandwidth, $B$ & 10 MHz & 500 MHz\\
        \hline
        Thermal noise, $N_0$ & -174 dBm/Hz & -174 dBm/Hz\\
        \hline
        Noise figure & 5 dB &0 dB \\
        \hline
        Shadowing variance, $\sigma_s^2$ & 9 dB & 12 dB\\
       \hline
        TX power, $P^{\mathrm{Tx}}$ & 46 dBm & 20 dBm\\
       \hline
        Antenna gain, $G^{\mathrm{Tx}}/G^{\mathrm{Rx}}$ & 17 dBi / 0 dBi & Fig.5\\ %\fig{fig:diagrams}\\
       \hline
        Radius, r & & 35m\\
       \hline
        Back-lobe gain & & -20dBi\\
       \hline
        Path-loss coefficient, $\eta_s$ & 3.76 & 2.5\\
       \hline
        Inter-cell distance & & $1.2\times r$\\
       \hline
        Reference distance, $d_{0,s}$ & $20.7\text{m}^{(1)}$  & $5$m \\
       \hline
        Beam number, $N_i$ & & $N_1=2$; $N_2=N_3=3$\\
       \hline
    \end{tabular}
    
    \begin{tablenotes}
    \small
    \item {\footnotesize $^{(1)}$ We use as a path loss model, $G_{i,j}^{\mathrm{ch.}}(\mathrm{dB}) = 128.1+37.6\mathrm{log}_{10}(d_{i,j})$, $d_{i,j}$ in Km from Table A.2.1.1.2-3 in \cite{TR36814}. Then, we compute the equivalent reference distance in meter for equation (3).}
    \end{tablenotes}
    \end{threeparttable}}
    %\end{minipage}%
\end{table}
    %\vfill
    %--------------------------------------------------------------------------------%
\begin{table}[!t]
    \centering
    %\begin{minipage}{.5\linewidth}
        \begin{threeparttable}
        \centering
        \caption{Training parameters}
        \label{tab:hyp-params}
        \renewcommand{\arraystretch}{1.3}
        \begin{tabular}{l||c}
           \hline
            Discount factor, $\gamma$ & 0.9\\
            \hline
            Time horizon, $T_e$ & 7000\\
            \hline
            Batch size, $|\mathcal{B}|$ & 32\\
            \hline
            CERTs memory size,  $|\mathcal{M}|$ & 500\\
            \hline
            $\epsilon$ (follows a negative Gompertz function$^{(1)}$) & $1\rightarrow0.1$ \\
            \hline
            Target network update frequency& 10\\
            \hline
            KL temperature, $\tau$ & 0.01\\
            \hline
            Number of Monte-Carlo simulations, N & 400\\
           \hline
        \end{tabular}
        
        \begin{tablenotes}
        \small
        \item {\footnotesize $^{(1)}\epsilon(t) = 1 - ae^{-b(t-c)}$, with $a=0.9$, $b=10^{-3}$, $c=800$.}
        \end{tablenotes}
        \end{threeparttable}
    %\end{minipage} 
\end{table}

To learn the user association policy, we use the DRQN described in \fig{fig:qnet}. This architecture comprises 2 multi-layers perceptron (MLP) of 32 hidden units, one RNN layer (a long short memory term - LSTM) layer with 64 memory cells followed by another 2 MLPs of 32 hidden units. The network then branches off in two MLPs of 16 hidden units to construct the duelling network. All layers use a rectifier linear unit (ReLU) except the final layer, which has a linear activation function. For the hysteretic learning, we set the base learning rate $\mu=0.001$ and $\alpha=1$, and then we optimize $\beta \in [0,1]$ to strike a balance between convergence speed and network sum-rate. The DRQNs are trained offline using an $\epsilon$-greedy policy. The hyper-parameters values summarized in Table \ref{tab:hyp-params} are selected via informal search. Finally, unless specified, all results are average over $N$ runs of Monte-Carlo simulations. At each run, UE positions are randomly reset. 

We evaluate the performance of the proposed solution and the related baselines using either the network sum-rate or the sum-rate ratio with respect to the brute force approach. Specifically, for these metrics, we compute the average and the standard deviation as follows:
\begin{IEEEeqnarray}{ll}
    \overline{R} &= \frac{1}{N}\sum_{n=1}^{N} \frac{1}{T_e}\sum_{t=1}^{T_e} R^{(n)}(t),\IEEEyesnumber\IEEEyessubnumber\\
    \sigma_R &= \sqrt{\frac{1}{N}\sum_{n=1}^{N} \left(\frac{1}{T_e}\sum_{t=1}^{T_e} R^{(n)}(t) - \overline{R}\right)^2},\IEEEyessubnumber
\end{IEEEeqnarray}
where $R^{(n)}(t)$ is either the sum-rate or the sum-rate ratio at the time step $t$ of run $n$.

\subsection{Complexity analysis}
In this subsection, we analyze both computational and signaling complexity of the proposed algorithm and compare it to the two baselines. Since our framework is based on deep Q-learning, a practical implementation completely conducts the learning offline as with the Vanilla DQN initially proposed for Atari games \cite{mnih2015humanlevel}, and then, it transfers to each UE the corresponding weights. In this scenario, UEs simply conduct the inference on their local states to find the optimal action, alleviating the computational and power burdens. That is to say, the computational complexity of the proposed framework during its execution is limited to the inference complexity of each local DQRN. Let $L_h$ be the size of hidden layers and $L_c$ the number of cells in the LSTM layer. Each DQRN has six inputs\footnote{Though for practical implementation, we encode the entry $a_j(t)$ in \fig{fig:qnet} as a one hot vector, thus, there are $5+\card{\mathcal{A}_j}$ inputs.}, thus the complexity is in the order of $\too{6L_h + 2L_h^2 + L_hL_c + 2L_h^2 + L_h(\card{\mathcal{A}_j})} \approx \too{6L_h^2 + L_hL_c}$. This is very straightforward compare to a naive algorithm, which may find the optimal solution of problem \eqref{eq:Obj}-\eqref{eq:C3} through an exhaustive search, which has a complexity $\too{N_sK(1+N_s)^{K}}$.
\begin{proof}
For UE $j$ there are $\card{\mathcal{A}_j}$ possible choices of BSs. The optimal association $\{i, ~\text{s.t.} ~x_{i,j}=1~\forall i\in\mathcal{A}\}$ is an element of $\underset{j\in\mathcal{U}}{\times}\mathcal{A}_j$. That is, for all UEs, there are $\prod_{j\in\mathcal{U}}\card{\mathcal{A}_j}$ possible combinations in which only some of them satisfy the constraint \eqref{eq:C2}. For each combination, checking if constraint (\ref{eq:C2}) is satisfied required $\too{\sum_{i\in\mathcal{A}} \card{\mathcal{U}_i}}$ iterations. In the worst case, when each UE can associate with any BS, $\card{\mathcal{A}_j}=N_s+1$. Hence, noting that ${\sum_{i\in\mathcal{A}} \card{\mathcal{U}_i} \leq N_sK}$, the complexity of running this naive algorithm will be therefore \[\too{\sum_{i\in\mathcal{A}} \card{\mathcal{U}_i}\prod_{j\in\mathcal{U}}\card{\mathcal{A}_j}} = \too{N_sK(1+N_s)^{K}}.\] 
\end{proof}

The complexity of both max-SNR and heuristic algorithms during execution is related to sorting the SNR values. Considering a \textit{quicksort} algorithm, this complexity in the worst case ($\card{\mathcal{A}_j}=N_s+1$) is around $\too{n\mathrm{log}(n)}$ for max-SNR and $\too{n+n\mathrm{log}(n)}$ for the heuristic algorithm\footnote{One pass to sort the SNR values and another to find the association.} where $n=K(1+N_s)$. However, the need to collect the SNR values globally is the most notable disadvantage of these centralized approaches. In terms of signaling overhead, compared to the existing standard (e.g. 5G), the additional complexity introduced by our framework is the broadcasting of the total network sum-rate. The rest of the information used by a UE to take a decision is already either measured by the UEs $\left(R_j(t), \mathrm{RSSI}_j, a_j(t)\right)$ or sent by its serving BS $\left(\mathrm{ACK}_j\right)$. Specifically, the number of messages exchanged in the sequence chart of \fig{fig:infoex} is a function of the UE's action $a_j(t)$. If $a_j(t)=0$, the association is set up in two messages with the MBS. Otherwise, four messages are required to connect to either a SBS or a MBS depending on the ACK signal. Overall, for each UE to connect to the serving BS, the system needs to exchange at most four messages. Then, two additional messages are required to get the total network sum-rate from the MBS. Therefore, at most six messages are needed to complete one training step.

\subsection{Convergence and effect of hysteretic parameter $\beta$}
\begin{figure}[!t]
    %----------------------     STATIC 5X5     -------------------------%
    \centering
    \subfloat[Loss function for different values of $\beta$ and $K=9$. For the sake of readability, a 20-sized moving average window is applied on plotted data.]{
         \includegraphics[scale=\figscale]{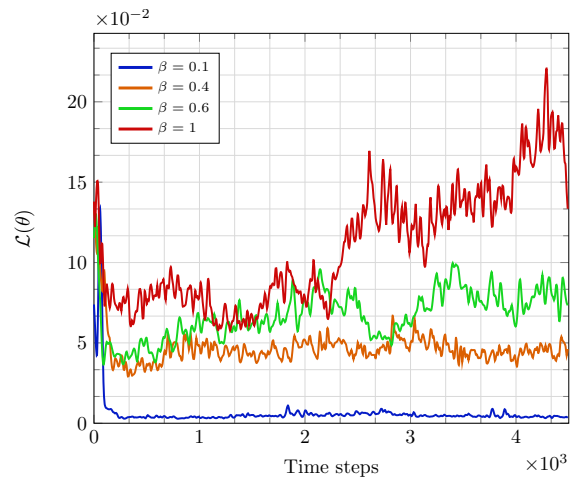}
        \label{fig:losses}
    }
    \hfill
    \subfloat[Sum-rate ratio and the associated variance between the proposed scheme and the optimal UE association for different values of $\beta$.]{
        \includegraphics[scale=\figscale]{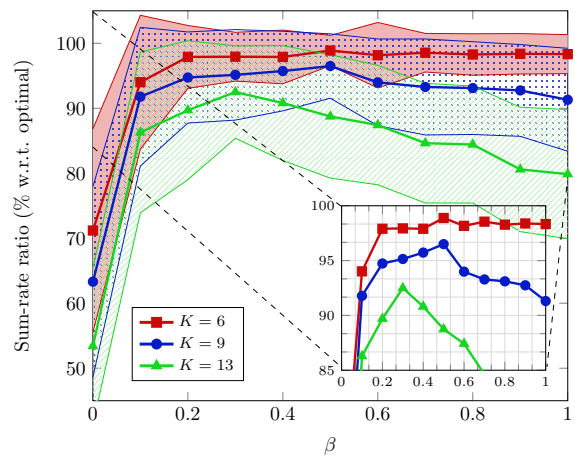}
        \label{fig:hysteretic}
    }
    \caption{Convergence speed and effect of the hysteretic parameter $\beta$, using diagram \texttt{diag 1}.}
    \label{fig:convergence}
\end{figure}

In this section, we study the impact of the hysteretic parameter $\beta$ on the performance of the proposed solution in terms of network sum-rate and convergence speed. Specifically, \fig{fig:losses} shows the evolution of the loss function during the training process for different values of $\beta$ and  \fig{fig:hysteretic} describes the sum-rate ratio of the proposed scheme with respect to the optimal solution as a function of $\beta$.

First, \figs{fig:losses}{fig:hysteretic} show that despite the few pieces of information available locally to each agent, they can successfully learn a user association policy that performs close to the optimal strategy in less than $5\cdot10^3$ iterations/associations if $\beta \leq 0.6$. In addition, \fig{fig:losses} shows that lowering $\beta$ increases the convergence speed of the algorithm. However, this also result in limited sum-rate performance. For instance, when $\beta=0$, the proposed scheme achieves only $70\%$ of the optimal performance (see \fig{fig:hysteretic}). This is because, from \eqref{hystereticweights}, we know that selecting very low values of $\beta$ makes the agents too optimistic i.e., they tend to neglect actions that produces negative TD errors. This leads agents to potentially select sub-optimal actions. 
 
In contrast, when $\beta=1$, the agents give equal importance to positive and negative TD errors, i.e., they become pessimist. In this setting, a UE may change its (optimal) strategy after taking an action that results in negative error, although this error is simply the result of the other agents' behaviors. These continuous changes limit the learning performance, and, in fact, \fig{fig:losses} shows that the loss function diverges for $\beta=1$.  Hence, there is a trade-off between convergence speed, successful coordination of the agents, and network sum-rate.
 
Moreover, as we can note from the zoom in \fig{fig:hysteretic}, if setting $\beta=0.5$ for $K=6$ UEs leads to the best performance, when increasing the number of UEs (i.e., $K=13$), the same value of $\beta$ results in drastic decrease of the sum-rate ratio performance, i.e., from $98\%$ to $88\%$. This result suggests that, depending on the number of UEs, there is an optimal hysteretic parameter $\beta$. Consequently, for the rest of the paper, we set $\beta=0.5$ for $K\in\{6,9\}$ and $\beta=0.3$ for $K=13$.
\vspace{-0.4cm}

\subsection{Impact of the collision cost on network performance}
\begin{figure}
    \centering
    \subfloat[]{
        \includegraphics[scale=0.8]{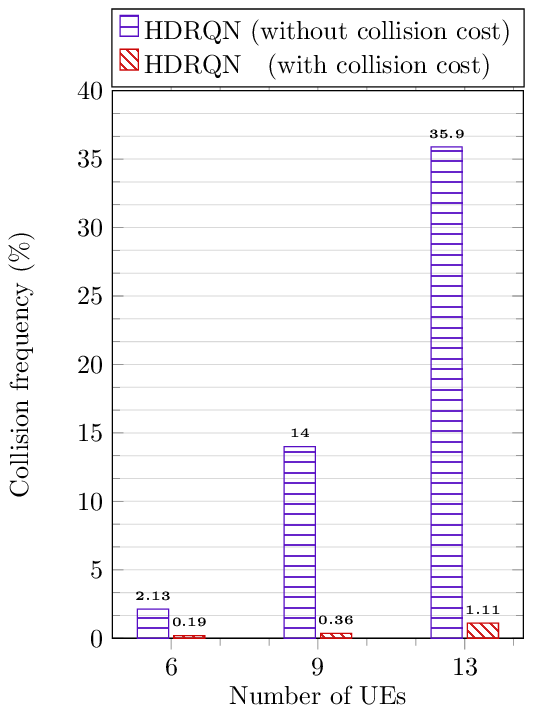}
        \label{fig:costcol1}
    } 
    \hfil
    \subfloat[]{
        %\hspace{-2.cm}
        \includegraphics[scale=0.8]{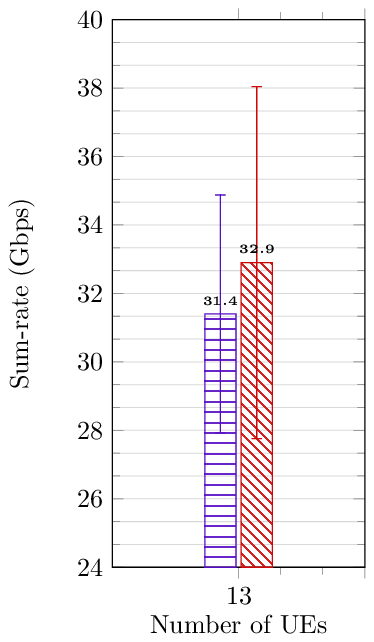}
        \label{fig:costcol2}
    }
    \caption{Impact of the collision cost on network performance in static scenario using diagram \texttt{diag 1}.}
    \label{fig:costcol}
\end{figure}
Here we assess the effectiveness of the collision cost in \eqref{eq:reward-def}, to limit the collision events. For this purpose, we consider a setting in which there is no collision cost. In this case, during the training phase, if a SBS receives a number of request greater than the ones that it can accept, it randomly chooses the serving UEs among the received requests; the remaining UEs are therefore associated with the MBS. \fig{fig:costcol1} shows the frequency of the collision event during the test phase. We can observe that the collision frequency increases with the number of UEs as the cell load increases. However, we can see that by introducing the collision penalty, we are able to significantly reduce the collision events up to $97\%$, which leads to an improvement of the overall network throughput\footnote{We have considered the case of $K=13$ UEs to highlight how, in networks with a large number of users, the collision events impact the network sum-rate.} by $4.7\%$ (see \fig{fig:costcol2}). This demonstrates that, with the proposed solution, UEs learn to distribute their association requests among the different BSs, balancing the cell load and maximizing the network sum-rate.

\subsection{Performance of the proposed algorithm in static scenario}
\begin{figure}
    %----------------------     STATIC 5X5     -------------------------%
    \centering
    \subfloat[Case 1: Using diagram \texttt{diag 1}]{
        \includegraphics[scale=\figscale]{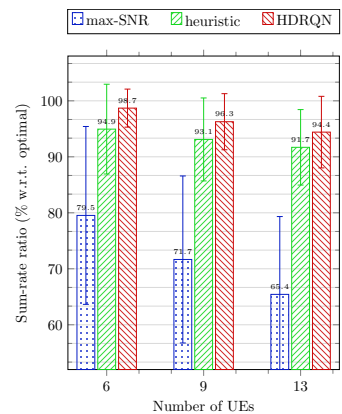}
        \label{fig:20x20norssi}
    } 
    \hfil
    \subfloat[Case 2: Using diagram \texttt{diag 3}]{
        \includegraphics[scale=\figscale]{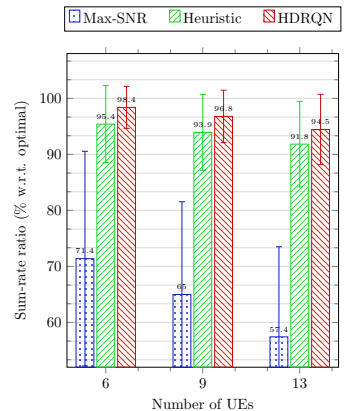}
        \label{fig:5x5norssi}
    }
    \caption{Performance comparison in static scenario using diagrams \texttt{diag 1} and \texttt{diag 3}.}
    \label{fig:static}
\end{figure}

We now compare the performance of the proposed user association solution with the one achieved by the two baselines in a static scenario where there is no fading (i.e., $\alpha_{i,j}=1$) and with full buffer traffic. Consequently, in \eqref{def-rate}, we set $D_j = +\infty, \forall j$ and disable the corresponding input in the DRQNs\footnote{To disable an input, we simply set the corresponding entry in $o(t)$ to zero.}. \figs{fig:20x20norssi}{fig:5x5norssi} show the performance of the different approaches compared to the optimal user association in terms of network sum-rate, using antenna diagrams \texttt{diag 1} and \texttt{diag 3} respectively. We first note that the sum-rate ratio performance of our solution as well as the heuristic approach barely changes between the two antenna diagrams (less than $0.5\%$ change), in contrast to the max-SNR algorithms, which does not consider interference. Specifically, when $K=13$ the performance of the max-SNR decreases by $12.2\%$ when switching from diagram \texttt{diag 1} to \texttt{diag 3}, which has lower directivity and thus results in a lower SINR. In addition, we note that, in average, our proposed scheme achieves up to $98.7\%$ of the optimal sum-rate, hence outperforming both the max-SNR and the heuristic approaches.
For example, when $K=6$, by using \texttt{diag 3}, the proposed solution exhibits a performance gain of $3.1\%$ and $37.8\%$ over the heuristic and the max-SNR algorithm, respectively.  

As soon as the number of UEs increases, the performance of our scheme slightly decreases. This is because ensuring coordination becomes more complex when the number of interacting agents increases. For instance, with \texttt{diag 3}, our solution only achieves $94.5\%$ of the optimal performance for $K=13$. However, it still outperforms the two baselines showing now a gain of $3\%$ and $64.6\%$ over the heuristic and the max-SNR approaches, respectively. Although the gain of the proposed solution over the heuristic scheme is small, one has to notice that our framework is distributed while the heuristic approach is centralized.

\subsection{Performance in dynamic scenarios}

\begin{figure*}[!t]
    %----------------------     FADING     -------------------------%
    \centering
    \subfloat[Case 1: Using diagram \texttt{diag 1}]{
         \includegraphics[scale=0.75]{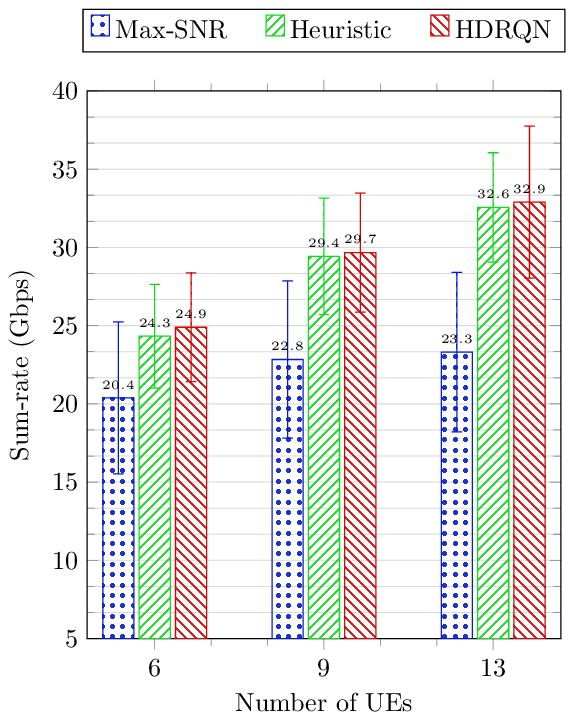}
        \label{fig:20x20fading}
    } 
    \hfil
    \subfloat[Case 2: Using diagram \texttt{diag 2}]{
         \includegraphics[scale=0.75]{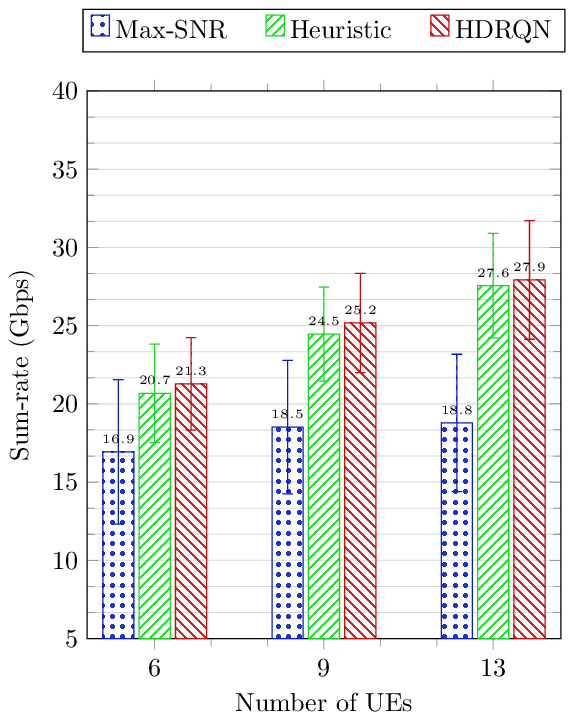}
        \label{fig:10x10fading}
    }
    \hfil
    \subfloat[Case 3: Using diagram \texttt{diag 3}]{
        \includegraphics[scale=0.75]{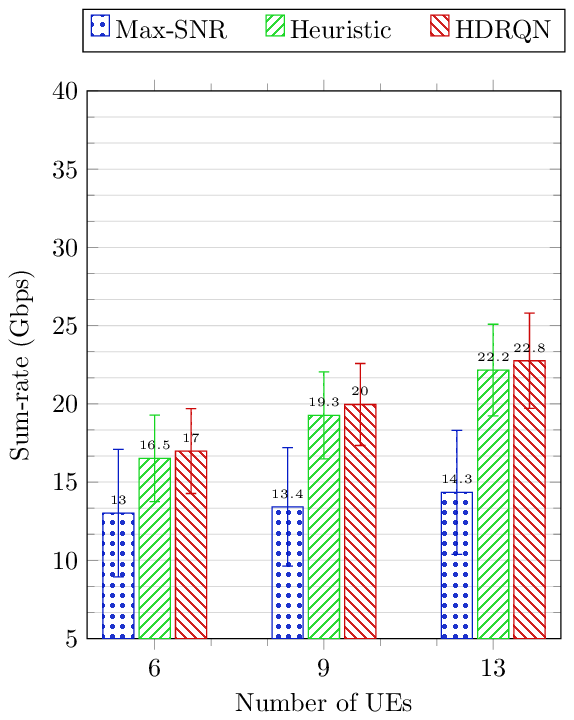}
        \label{fig:5x5fading}
    }       
    \caption{Performance comparison when considering only dynamic channels with fast fading.}
    \label{fig:fading}
\end{figure*}

\begin{figure*}[!t]
    %----------------------     TRAFFIC     -------------------------%
    \centering
    \subfloat[Case 1: Using diagram \texttt{diag 1}]{
        \includegraphics[scale=\figscale]{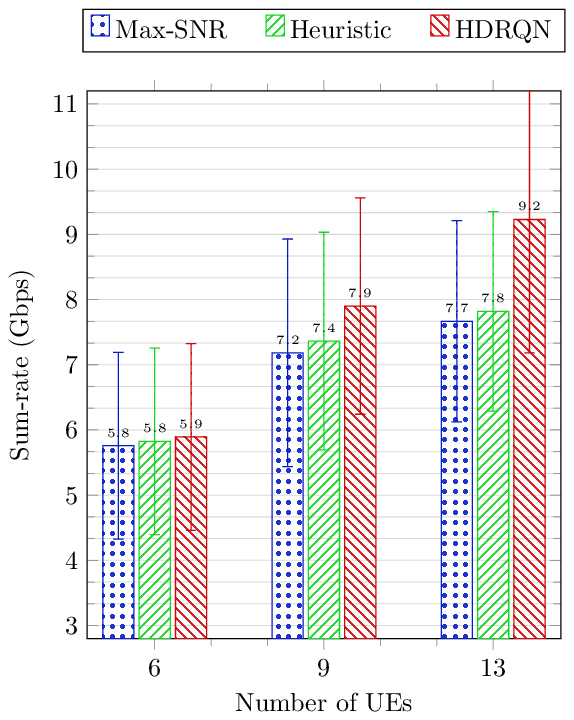}
        \label{fig:20x20traffic}
    }       
    \hfil
    \subfloat[Case 2: Using diagram \texttt{diag 2}]{
        \includegraphics[scale=\figscale]{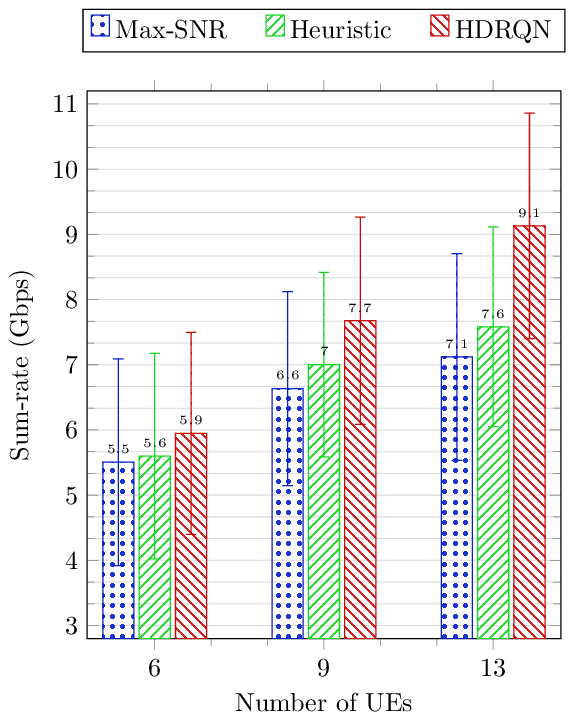}
        \label{fig:10x10traffic}
    }
    \hfil
    \subfloat[Case 3: Using diagram \texttt{diag 3}]{
        \includegraphics[scale=\figscale]{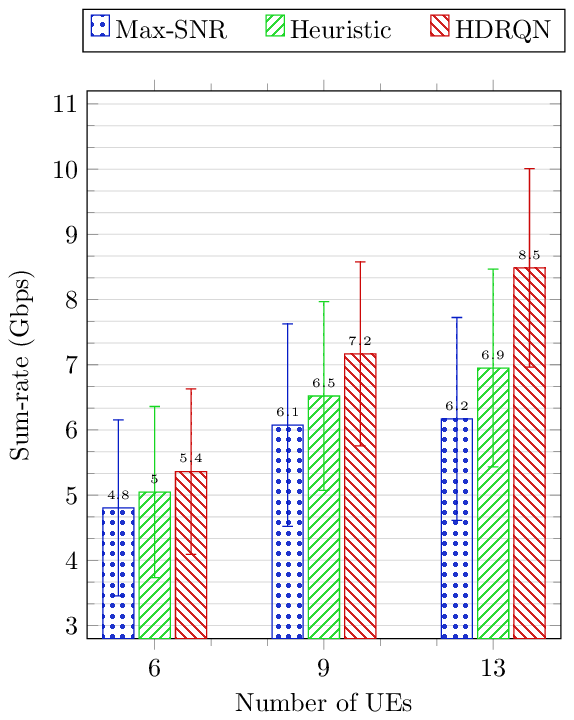}
        \label{fig:5x5traffic}
    } 
    \caption{Performance comparison when considering only dynamic traffic.}
    \label{fig:traffic}
\end{figure*}

\begin{figure*}[!t]
    %----------------------  FADING TRAFFIC     -------------------------%
    \centering
    \subfloat[Case 1: Using diagram \texttt{diag 1}]{
        \includegraphics[scale=\figscale]{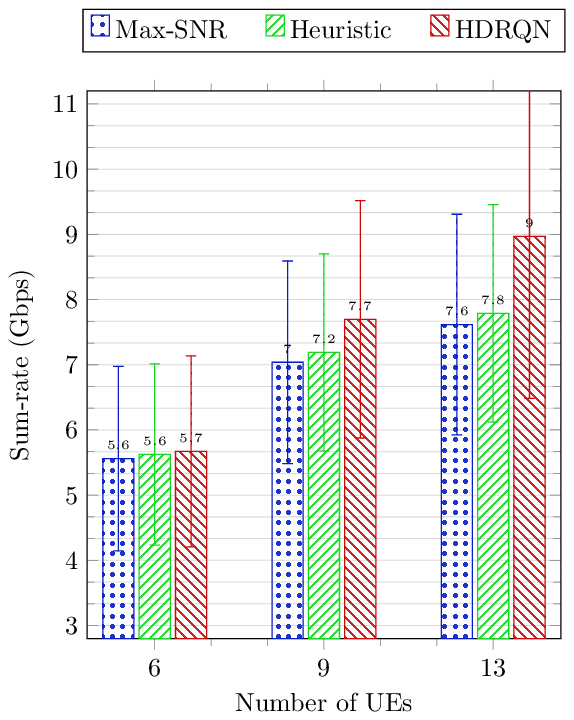}
        \label{fig:20x20fading-traffic}
    }     
    \hfil
    \subfloat[Case 2: Using diagram \texttt{diag 2}]{
        \includegraphics[scale=\figscale]{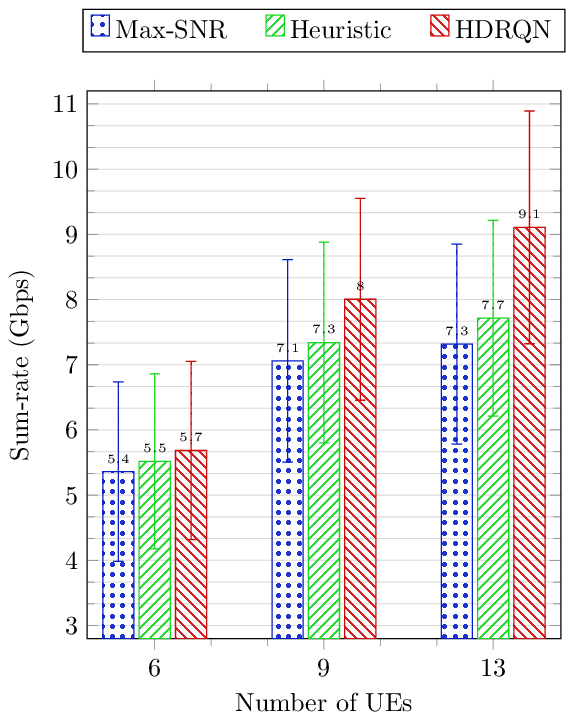}
        \label{fig:10x10fading-traffic}
    }
    \hfil
    \subfloat[Case 3: Using diagram \texttt{diag 3}]{
        \includegraphics[scale=\figscale]{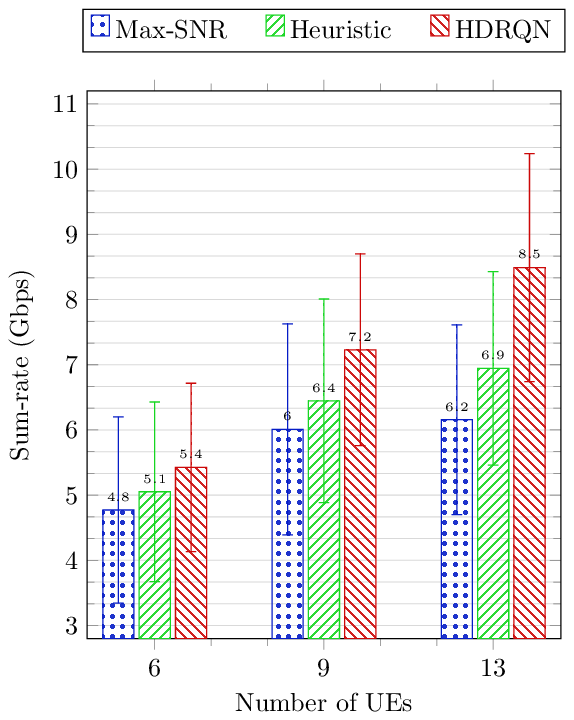}
        \label{fig:5x5fading-traffic}
    }   
    \caption{Performance comparison when considering both dynamic channels with fast fading and dynamic traffic.}
    \label{fig:fading-traffic}
\end{figure*}

We now evaluate the performance of the proposed scheme in dynamic environments and considering the three different antenna diagrams in \fig{fig:diagrams}. For this purpose, we define three cases: 1) dynamic channels with small scale fading and full buffer traffic,  2) static channels with dynamic traffic, 3) dynamic channels and dynamic traffic. As the optimal user association obtained via exhaustive search requires extensive computation, in the following, unless otherwise stated, we compare only the performance of the proposed scheme with the aforementioned two baselines. To achieve fair comparison, every time step, we recompute the association solution of the two baselines as this may change due to environment dynamics.

Before discussing the three aforementioned cases, we can highlight from \figss{fig:fading}{fig:traffic}{fig:fading-traffic} that, as expected, the network sum-rate decreases as the antenna diagrams become less and less directive (from \texttt{diag 1} to \texttt{diag 3}).
\newline
\subsubsection{\textbf{Dynamic channel with small scale fading}}
In this scenario, we have full buffer traffic, $D_j = +\infty, \forall j$ and dynamic channels with Nakagami small scale fading, characterized by a scale factor $m=3$ \cite{chevillon2018}. \fig{fig:fading} plots the sum-rate achieved by the different algorithms for a different number of UEs. We remark that our distributed solution performs better than the two centralized baselines. Specifically, when the number of UEs is equal to 9, the HDRQN improves the network sum-rate by about $1\%$ and $30.3\%$ when using \texttt{diag 1},  $2.8\%$ and $36.2\%$ when using \texttt{diag 2}, and $3.6\%$ and $49.2\%$ when using \texttt{diag 3}, compared respectively to the heuristic and max-SNR schemes. As in the static case, the gain with respect to the heuristic is limited when considering only the fast fading effect. Also, the gap between our scheme and the two baselines decreases when the antennas are more directive, which is due to the smaller interference perceived at the UE side. In fact, the two baselines perform better in limited interference scenarios.
\newline
\subsubsection{\textbf{UE traffic}}
Now we consider static channels (i.e., $\alpha_{i,j}=1,~\forall i,j$) and dynamic traffic. For each UE, the intensity of its traffic Poisson distribution is uniformly chosen between $[0, 2]$ ~\text{Gbps} at the beginning of each Monte Carlo run. \fig{fig:traffic} shows the total network throughput reached by the different algorithms. In contrast to the previous case shown in \fig{fig:fading},  we can remark that the our algorithm yields a large performance gain over the two benchmarks. For instance, for $K=13$ UEs, the proposed solution improves the sum-rate by $19.4\%$ and $18\%$ when using \texttt{diag 1}, $19.7\%$ and $28.2\%$ when using \texttt{diag 2}, $23.2\%$ and $37.1\%$ with \texttt{diag 3}, compared to heuristic algorithm and the max-SNR algorithms, respectively. 
\newline

\subsubsection{\textbf{UE traffic and Fast fading channels}}

Finally, we evaluate the performance of our framework considering both fast fading and UE traffic. \fig{fig:fading-traffic} exhibits similar trend as the case with dynamic traffic only. Overall, as expected, the effect of the traffic variations on the rate (see \eqref{def-rate}) is larger than the one related to the fast fading, which leads to small variations on the user perceived SINR (see \eqref{SINR}).

\subsection{Performance  with respect to change in service requests}
\begin{figure}[!t]
    \centering
    \includegraphics[scale=0.75]{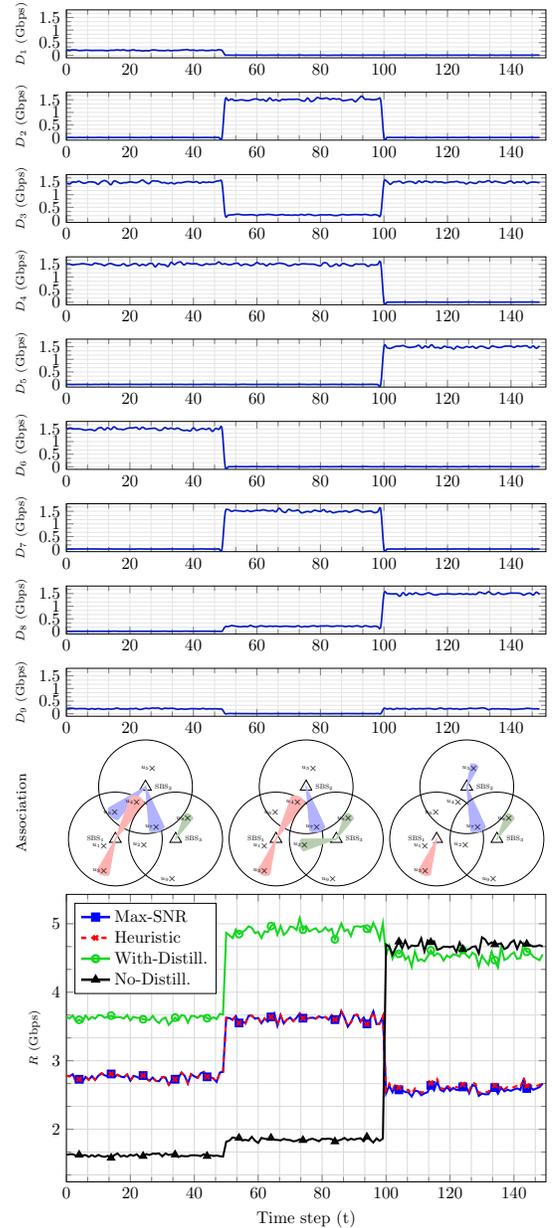}
    \caption{Dynamic behavior of the proposed adaptive user association scheme. We set the loss temperature to $\tau=0.01$ via informal search.}
    \label{fig:adaptive}
\end{figure}
Finally, we show the capacity of our scheme to adapt the association policy with respect to time-varying service requests in the network. As the service request (and its corresponding data rate) at each UE change during time, the user association has to adapt to keep the network performance optimized i.e., balancing the cell load. To achieve this property, we use the distillation mechanism described in Section \ref{section:distillation}. 

Let us consider three services rate requirements denoted as \texttt{service 1}, \texttt{service 2}, and \texttt{service 3}, corresponding respectively to an average data rate demand of $D_{\mathrm{s1}}=5~\mathrm{Mbps}$, $D_{\mathrm{s2}}=200~\mathrm{Mbps}$, and $D_{\mathrm{s3}}=1.5~\mathrm{Gbps}$. \texttt{service 1} may be related to web browsing or voice call services, \texttt{service 2} to online video streaming, and \texttt{service 3} to augmented reality or virtual reality applications. 
In the following, we focus on three time periods during which the UEs randomly change their service requests and we apply the distillation procedure (Algorithm \ref{algo:distillation} with $P=3$). 
\fig{fig:adaptive} shows a sample of the performance of the proposed HDRQN with and without distillation. Specifically, the agent policies without distillation is obtained through a single training phase over the three time periods. The upper part of this figure highlights the data rate changes for each of the 9 UEs in the network. The middle part of the figure describes the corresponding user association\footnote{Note that the UEs not served by a mmWave beam are receiving data through the MBS.}. Finally, the lower part shows the evolution of the network sum-rate. Overall, \fig{fig:adaptive} shows that the proposed algorithm using distillation mechanism can effectively adapt the user association to service request dynamics thus outperforming the two baselines. For example, we see that UE 4 is served by SBS 1 in the first two time intervals, when it is requiring \texttt{service 3}; in contrast, in the last interval, when it demands for \texttt{service 1}, which is characterized by a lower data rate request, its access is provided by the MBS. Meanwhile, in the last interval, UE 5 asks for \texttt{service 3}; therefore, it hands off from the MBS to SBS 2, which can satisfy its demand for higher data rate. Moreover, we can highlight that, in the absence of distillation, the proposed solution shows poor performance during the first two time periods. This is due to the forgetting effect inherent to neural networks training: at the end of the third period, the agents have forgotten what they have learned in the first two periods. The resulting policy is therefore only appraised to handle the last service for which it exhibits the best performance.

\section{Conclusion}\label{conclusion}
In this paper, we have investigated the problem of user association in dense millimeter-wave networks with multiple radio access technologies. To deal with network dynamics,  we have presented a novel and distributed approach based on deep reinforcement learning. With the proposed MARL algorithm, agent decisions are based on partial and local observations, which limits the signaling overhead and reduces the computational complexity with respect to centralized approaches. In addition, by integrating a \textit{distillation} procedure, we make our proposed solution robust to fast changes in the statistics of the environment dynamics (interference, fading, traffic). Our analysis shows that, in cases of full buffer traffic, the proposed scheme achieves up to $98.7\%$ of the optimal performance obtained through exhaustive search. When considering dynamic fading, the proposed solution outperforms centralized baselines, which require to continuously recompute the user association and hence, are characterized by excessive complexity. In addition, our approach results in large sum-rate gains when we consider also dynamic traffic. In fact, numerical results show that, in this context, the proposed algorithm can achieve nearly $40\%$ of performance gain with respect to baseline solutions from the literature.

In future work, we will include user mobility as an additional case of environment dynamics, and we will investigate how to include SBS in the decision process, for instance to select which users to serve when multiple requests are received simultaneously.


\begin{thebibliography}{10}

\bibitem{sana2019UA}
M.~{Sana}, A.~{De Domenico}, and E.~{Calvanese Strinati}, ``{Multi-Agent Deep
  Reinforcement Learning based User Association for Dense mmWave Networks},''
  {\em in IEEE Global Communications Conference (GLOBECOM)}, pp.~1--6, Dec
  2019.

\bibitem{TR38.913}
\relax 3GPP TR~38.913, ``{5G; Study on Scenarios and Requirements for Next
  Generation Access Technologies (Release 15)},'' Sept 2018.

\bibitem{DeDomenico2017}
A.~{De Domenico}, R.~{Gerzaguet}, N.~{Cassiau}, A.~{Clemente}, R.~{D'Errico},
  C.~{Dehos}, J.~L. {Gonzalez}, D.~{Ktenas}, L.~{Manat}, V.~{Savin}, and
  A.~{Siligaris}, ``{Making 5G Millimeter-Wave Communications a Reality
  [Industry Perspectives]},'' {\em IEEE Wireless Communications}, vol.~24,
  pp.~4--9, Aug 2017.

\bibitem{emilio2019}
E.~{Calvanese Strinati}, S.~{Barbarossa}, J.~L. {Gonzalez-Jimenez},
  D.~{Ktenas}, N.~{Cassiau}, L.~{Maret}, and C.~{Dehos}, ``{6G: The Next
  Frontier: From Holographic Messaging to Artificial Intelligence Using
  Subterahertz and Visible Light Communication},'' {\em IEEE Vehicular
  Technology Magazine}, vol.~14, pp.~42--50, Sep. 2019.

\bibitem{athanasiou}
G.~{Athanasiou}, P.~C. {Weeraddana}, C.~{Fischione}, and L.~{Tassiulas},
  ``{Optimizing Client Association for Load Balancing and Fairness in
  Millimeter-Wave Wireless Networks},'' {\em IEEE/ACM Transactions on
  Networking}, vol.~23, pp.~836--850, June 2015.

\bibitem{lui}
Y.~{Liu}, X.~{Fang}, M.~{Xiao}, and S.~{Mumtaz}, ``{Decentralized Beam Pair
  Selection in Multi-Beam Millimeter-Wave Networks},'' {\em IEEE Transactions
  on Communications}, vol.~66, pp.~2722--2737, June 2018.

\bibitem{mao}
Q.~{Mao}, F.~{Hu}, and Q.~{Hao}, ``{Deep Learning for Intelligent Wireless
  Networks: A Comprehensive Survey},'' {\em IEEE Communications Surveys
  Tutorials}, vol.~20, pp.~2595--2621, Fourthquarter 2018.

\bibitem{busoniu2008}
L.~{Busoniu}, R.~{Babu${\hat s}$ka}, and B.~D. {Schutter}, ``{A Comprehensive
  Survey of Multiagent Reinforcement Learning},'' {\em IEEE Transactions on
  Systems, Man, and Cybernetics, Part C (Applications and Reviews)}, vol.~38,
  pp.~156--172, March 2008.

\bibitem{zhou}
P.~{Zhou}, X.~{Fang}, X.~{Wang}, Y.~{Long}, R.~{He}, and X.~{Han}, ``{Deep
  Learning-Based Beam Management and Interference Coordination in Dense mmWave
  Networks},'' {\em IEEE Transactions on Vehicular Technology}, vol.~68,
  pp.~592--603, Jan 2019.

\bibitem{deng2019deep}
Y.~Deng, ``{Deep Learning on Mobile Devices: a Review},'' in {\em Mobile
  Multimedia/Image Processing, Security, and Applications}, vol.~10993,
  p.~109930A, International Society for Optics and Photonics, 2019.

\bibitem{lee2019device}
J.~Lee, N.~Chirkov, E.~Ignasheva, Y.~Pisarchyk, M.~Shieh, F.~Riccardi,
  R.~Sarokin, A.~Kulik, and M.~Grundmann, ``{On-device Neural Net Inference
  with Mobile GPUs},'' {\em arXiv preprint arXiv:1907.01989}, 2019.

\bibitem{zhao}
N.~{Zhao}, Y.~{Liang}, D.~{Niyato}, Y.~{Pei}, and Y.~{Jiang}, ``{Deep
  Reinforcement Learning for User Association and Resource Allocation in
  Heterogeneous Networks},'' in {\em Proc. IEEE Global Communications
  Conference (GLOBECOM)}, pp.~1--6, Dec 2018.

\bibitem{sana2019HO}
M.~{Sana}, A.~{De Domenico}, and E.~{Calvanese Strinati}, ``{Multi-Agent Deep
  Reinforcement Learning for Distributed Handover Management in Dense mmWave
  Networks},'' {\em IEEE International Conference on Acoustics, Speech, and
  Signal Processing (ICASSP)}, May 2020.

\bibitem{Li2017}
Y.~{Li}, J.~G. {Andrews}, F.~{Baccelli}, T.~D. {Novlan}, and C.~J. {Zhang},
  ``{Design and Analysis of Initial Access in Millimeter Wave Cellular
  Networks},'' {\em IEEE Transactions on Wireless Communications}, vol.~16,
  pp.~6409--6425, Oct 2017.

\bibitem{chevillon2018}
R.~{Chevillon}, G.~{Andrieux}, R.~{Négrier}, and J.~{Diouris}, ``{Spectral and
  Energy Efficiency Analysis of mmWave Communications With Channel Inversion in
  Outband D2D Network},'' {\em IEEE Access}, vol.~6, pp.~72104--72116, 2018.

\bibitem{Rappaport2017}
T.~S. {Rappaport}, Y.~{Xing}, G.~R. {MacCartney}, A.~F. {Molisch},
  E.~{Mellios}, and J.~{Zhang}, ``{Overview of Millimeter Wave Communications
  for Fifth-Generation (5G) Wireless Networks With a Focus on Propagation
  Models},'' {\em IEEE Transactions on Antennas and Propagation}, vol.~65,
  no.~12, pp.~6213--6230, 2017.

\bibitem{Domenico2013BackHaul}
A.~{De Domenico}, V.~{Savin}, and D.~{Ktenas}, ``{A Backhaul-Aware Cell
  Selection Algorithm for Heterogeneous Cellular Networks},'' in {\em 2013 IEEE
  24th Annual International Symposium on Personal, Indoor, and Mobile Radio
  Communications (PIMRC)}, pp.~1688--1693, Sep. 2013.

\bibitem{Sapountzis2017}
N.~{Sapountzis}, T.~{Spyropoulos}, N.~{Nikaein}, and U.~{Salim}, ``{User
  Association in HetNets: Impact of Traffic Differentiation and Backhaul
  Limitations},'' {\em IEEE/ACM Transactions on Networking}, vol.~25,
  pp.~3396--3410, Dec 2017.

\bibitem{shen2018}
K.~{Shen} and W.~{Yu}, ``{Fractional Programming for Communication
  Systems—Part I: Power Control and Beamforming},'' {\em IEEE Transactions on
  Signal Processing}, vol.~66, no.~10, pp.~2616--2630, 2018.

\bibitem{bucsoniu2010multi}
L.~Bu{\c{s}}oniu, R.~Babu{\v{s}}ka, and B.~De~Schutter, ``{Multi-agent
  Reinforcement Learning: An Overview},'' in {\em Innovations in multi-agent
  systems and applications-1}, pp.~183--221, Springer, 2010.

\bibitem{mnih2015humanlevel}
V.~Mnih, K.~Kavukcuoglu, D.~Silver, A.~A. Rusu, J.~Veness, M.~G. Bellemare,
  A.~Graves, M.~Riedmiller, A.~K. Fidjeland, G.~Ostrovski, S.~Petersen,
  C.~Beattie, A.~Sadik, I.~Antonoglou, H.~King, D.~Kumaran, D.~Wierstra,
  S.~Legg, and D.~Hassabis, ``{Human-level Control through Deep Reinforcement
  Learning},'' {\em Nature}, vol.~518, pp.~529--533, Feb. 2015.

\bibitem{wang2015dueling}
Z.~Wang, T.~Schaul, M.~Hessel, H.~Hasselt, M.~Lanctot, and N.~Freitas,
  ``{Dueling Network Architectures for Deep Reinforcement Learning},'' in {\em
  Proc. International Conference on Machine Learning (PMLR)}, vol.~48,
  pp.~1995--2003, Jun 2016.

\bibitem{matignon2012independent}
L.~Matignon, G.~J. Laurent, and N.~Le~Fort-Piat, ``{Independent Reinforcement
  Learners in Cooperative Markov Games: a Survey Regarding Coordination
  Problems},'' {\em The Knowledge Engineering Review}, vol.~27, no.~1,
  pp.~1--31, 2012.

\bibitem{naparstek}
O.~Naparstek and K.~Cohen, ``{Deep Multi-User Reinforcement Learning for
  Distributed Dynamic Spectrum Access},'' {\em IEEE Transactions on Wireless
  Communications}, vol.~18, pp.~310--323, Jan 2019.

\bibitem{omidshafiei2017}
S.~Omidshafiei, J.~Pazis, C.~Amato, J.~P. How, and J.~Vian, ``{Deep
  Decentralized Multi-task Multi-Agent Reinforcement Learning under Partial
  Observability},'' in {\em Proc. International Conference on Machine Learning
  (ICML)}, vol.~70, pp.~2681--2690, PMLR, 06--11 Aug 2017.

\bibitem{matignon2007hysteretic}
L.~Matignon, G.~J. Laurent, and N.~Le~Fort-Piat, ``{Hysteretic Q-Learning: An
  Algorithm for Decentralized Reinforcement Learning in Cooperative Multi-agent
  Teams},'' in {\em Proc. International Conference on Intelligent Robots and
  Systems (IEEE/RSJ)}, pp.~64--69, 2007.

\bibitem{hausknecht2015deep}
M.~Hausknecht and P.~Stone, ``{Deep Recurrent Q-Learning for Partially
  Observable MDPs},'' in {\em AAAI Fall Symposium on Sequential Decision Making
  for Intelligent Agents (AAAI-SDMIA15)}, November 2015.

\bibitem{rusu2015policy}
A.~A. Rusu, S.~G. Colmenarejo, {\c{C}}.~G{\"{u}}l{\c{c}}ehre, G.~Desjardins,
  J.~Kirkpatrick, R.~Pascanu, V.~Mnih, K.~Kavukcuoglu, and R.~Hadsell,
  ``{Policy Distillation},'' in {\em Proc. International Conference on Learning
  Representations (ICLR)}, May 2016.

\bibitem{mailloux2017phased}
R.~J. Mailloux, {\em {Phased Array Antenna Handbook}}.
\newblock Norwood, MA, USA: Artech House, Inc., 3rd~ed., 2017.

\bibitem{TR36872}
\relax 3GPP TR~36.872, ``{Small cell enhancements for E-UTRA and E-UTRAN -
  Physical layer aspects (Release 12)},'' Dec 2013.

\bibitem{TR36814}
\relax 3GPP TR~36.814, ``{Evolved Universal Terrestrial Radio Access (E-UTRA) -
  Further advancements for E-UTRA physical layer aspects (Release 9)},'' Mars
  2017.

\end{thebibliography}
\end{document}